\documentclass[fleqn,usenatbib]{mnras}

\usepackage{newtxtext,newtxmath}
\usepackage{fix-cm}

\usepackage[T1]{fontenc}

\DeclareRobustCommand{\VAN}[3]{#2}
\let\VANthebibliography\thebibliography
\def\thebibliography{\DeclareRobustCommand{\VAN}[3]{##3}\VANthebibliography}

\usepackage{graphicx}	
\usepackage{amsmath}	

\usepackage{amssymb}	

\title[WALLABY pilot survey: NGC 4532 / DDO 137]{WALLABY Pilot Survey: the extensive interaction of NGC 4532 and DDO 137 with the Virgo cluster}

\author[L. Staveley-Smith et al.]
{\parbox{\textwidth}{L. Staveley-Smith,$^{1,2}$\thanks{E-mail: Lister.Staveley-Smith@uwa.edu.au} 
K. Bekki,$^{1}$ 
A. Boselli,$^{3}$ 
L. Cortese,$^{1,2}$ 
N. Deg,$^{4}$ 
B.-Q. For,$^{1,2}$ 
K. Lee-Waddell,$^{1,5,6}$ 
{T.~O’Beirne},$^{7,6,8}$ 
M.E. Putman,$^9$ 
C. Sinnott,$^{1}$ 
J. Wang,$^{10}$ 
T. Westmeier,$^{1,2}$ 
O.I. Wong,$^{6,1}$
B. Catinella,$^{1,2}$
H.~Dénes,$^{11}$
J. Rhee,$^{1}$
L. Shao,$^{12}$
A.X. Shen,$^{6}$ and
K. Spekkens$^{4}$ 
}
 \vspace{0.4cm}\\
\parbox{\textwidth}{
$^{1}$International Centre for Radio Astronomy Research (ICRAR), The University of Western Australia, 35 Stirling Highway, Crawley, WA 6009, Australia\\
$^{2}$ARC Centre of Excellence for All Sky Astrophysics in 3 Dimensions (ASTRO 3D), Australia\\
$^3$Aix Marseille Universit\'{e} , CNRS, CNES, LAM, 13013 Marseille, France\\
$^4$Department of Physics, Engineering Physics, and Astronomy, Queen’s University, Kingston ON K7L 3N6, Canada\\
$^5$International Centre for Radio Astronomy Research (ICRAR), Curtin University, Bentley, WA 6102, Australia\\
$^6$CSIRO Space \& Astronomy, PO Box 1130, Bentley WA 6102, Australia\\
$^7$Centre for Astrophysics and Supercomputing, Swinburne University of Technology, Hawthorn, Victoria 3122, Australia\\
$^8$European Southern Observatory, Karl-Schwarzschildstrasse 2, D-85748 Garching bei M\"unchen, Germany\\
$^9$Department of Astronomy, Columbia University, New York, NY 10027, USA\\
$^{10}$Kavli Institute for Astronomy and Astrophysics, Peking University, Beijing 100871, China\\
$^{11}$School of Physical Sciences and Nanotechnology, Yachay Tech University, Hacienda San José S/N, 100119, Urcuquí, Ecuador\\
$^{12}$National Astronomical Observatories, Chinese Academy of Sciences, Beijing 100101, China
}}

\date{Accepted 2025 August 27. Received 2025 August 26; in original form 2025 April 13}

\pubyear{2025}

\begin{document}
\label{firstpage}
\pagerange{\pageref{firstpage}--\pageref{lastpage}}
\maketitle

\begin{abstract}
As part of the pilot survey of the Widefield ASKAP L-band Legacy All-sky Survey (WALLABY), high-resolution neutral atomic hydrogen (HI) observations of the dwarf galaxy pair NGC 4532/DDO 137 (WALLABY J123424+062511) have revealed a huge {(48 kpc)} bridge of gas between the two galaxies, as well as numerous arms and clouds which connect with the even longer (0.5 Mpc) tail of gas previously discovered with the Arecibo telescope.    Our modelling suggests that a combination of ram pressure and tidal forces are responsible for the nature of the system. Although the pair lies well outside of the virial radius of the Virgo cluster, ram pressure due to infall through an extensive envelope of hot gas around the cluster is most likely responsible for the HI tail. Over a timescale of 1 Gyr, the predicted electron density ($1.2\times 10^{-5}$ cm$^{-3}$) and infall velocity (880 km s$^{-1}$) are probably sufficient to explain the extensive stripping from the common gaseous envelope of NGC 4532/DDO 137. The ongoing tidal interaction with the Virgo cluster appears to have prevented a rapid merger of the binary pair, with the mutual tidal interaction between the galaxy pair being responsible for raising gas from the outer parts of the galaxy potential wells into the HI bridge and common envelope. The NGC 4532/DDO 137 system mirrors many of the physical features of the Magellanic System, and may lead to a better understanding of that system, as well as casting more light on the relative importance of interaction mechanisms in the outskirts of dynamically young galaxy clusters such as Virgo.
\end{abstract}


\section{INTRODUCTION }
\label{sec:intro}
The formation and evolution of galaxies in the Universe involves the collapse and merging of multiple low-mass systems, with the rate of evolution being directly related to the local density and velocity dispersion. Galaxies in clusters evolve more rapidly than galaxies in filaments and in the low-density regions, {but there seems to be no sharp transition  at the cluster boundary} -- rather, there appears to be a smooth transition in parameters such as star formation rate, morphology  and gas content \citep{2021PASA...38...35C,2022A&A...657A...9C,2022A&ARv..30....3B} which may reflect the increase in the halo mass within which galaxies reside \citep{2019MNRAS.483.2851S}. However, a number of competing factors affect the evolution of galaxies, such as the changing ratio of major and minor interaction rates, the rate of high-velocity fly-bys, the role of ram-pressure stripping in dense intergalactic environments, and the effect of interactions on triggering feedback from star and black hole formation.

Since many of these evolutionary processes can be studied most conveniently in and around overdense regions, the pilot phase of the Widefield ASKAP L-band Legacy All-sky Survey (WALLABY) was designed to target groups, supergroups and clusters in the local Universe. The full WALLABY survey \citep{2020Ap&SS.365..118K} is an large area survey which is imaging the local Universe ($z<0.1$) in the 21-cm line of neutral hydrogen, and is very sensitive to the kinematic and morphological disturbances which result from interactions between gas-rich galaxies in the early stages of interaction before they lose the bulk of their HI mass through molecular gas formation, star formation, or photoevaporation. The pre-pilot and pilot surveys \citep{2022PASA...39...58W,2024PASA...41...88M} were designed as an end-to-end test of the ASKAP telescope \citep{Hotan2021}, the data reduction pipeline \citep{2020ASPC..522..469W}, the source-detection strategy \citep{2015MNRAS.448.1922S,2021MNRAS.506.3962W}, and kinematic modelling \citep{2022PASA...39...59D} using similar observing parameters (strategy, integration time, mosaicking, processing) as the full survey, but to include specific target such as the NGC 4636, NGC 5044 and NGC 4808 groups, the Eridanus supergroup, the Hydra and Norma clusters, and the Vela supercluster. Due to the large field (one or more 30 sq.deg tiles around each target) and redshift depth ($z < 0.1$) of the pilot ASKAP observations, an extensive region around each of the groups/clusters was also included. 

With a view to better understanding the nature and evolution of galaxies in different environments, extensive analyses of the above data has already been undertaken \citep{2021MNRAS.505.1891R,2021MNRAS.507.2300F,2021MNRAS.507.2905W,2021MNRAS.507.2949M,2021ApJ...915...70W,2022MNRAS.510.1716R,2022ApJ...927...66W,2023MNRAS.519..318K,2023MNRAS.519.4589C,2023MNRAS.521.1502H,2023PASA...40...32R,2023MNRAS.525.4663D,2023MNRAS.526.3130F,2024MNRAS.528.4010O,2024MNRAS.533..925M,2024ApJ...976..159D,2024arXiv241022406H}. During these investigations, the interacting galaxy pair NGC 4532/DDO 137 was detected. It was classified as a complex system and given the combined designation WALLABY J123424+062511. This system lies in the northern part of the NGC 4636 field, six degrees (1.7 Mpc in projection) south of the Virgo cluster (centred on M87), and 1.9 deg south of M49. It has been the subject of several previous Arecibo and Jansky Very Large Array (VLA) investigations \citep{1992ApJ...388L...5H,1993AJ....106...39H,1999AJ....117..811H,2008ApJ...682L..85K,2009AJ....138.1741C,2016MNRAS.459.1827P}. However, the full extent of the bridge between the two galaxies was not previously appreciated, so the new observations present an opportunity to re-visit formation scenarios, particularly in the light of our increased understanding of the relative role of tidal and ram-pressure forces in different situations, and our increasing capability to detect hot cluster gas  with new telescope surveys such as with eROSITA \citep{2021A&A...647A...1P}.

NGC 4532 and DDO 137 (the latter is also known as UGC 7739, Holmberg VII and VCC 1581) are a pair of irregular/Magellanic-type galaxies lying in the Virgo cluster southern extension, one of a number of filaments which lead to the Virgo cluster \citep{2021ApJ...906...68L}. Their distance is probably close to the Virgo cluster distance of 16.5 Mpc \citep{2007ApJ...655..144M,2024ApJ...966..145C}. 
The mean distance for all 20 redshift-independent distances for NGC 4532 in the NASA/IPAC Extragalactic Database (NED) \citep{2017AJ....153...37S} is 13.8 Mpc. These estimates are all based on the Tully-Fisher \citep[TF;][]{1977A&A....54..661T} relation , which has considerable errors for low-luminosity galaxies. Unfortunately, no `Tip of the Red Giant Branch' \citep[TRGB;][]{2018ApJ...858...62K} or other accurate redshift-independent distances are available. The heliocentric velocity of NGC 4532 is around 2011 km~s$^{-1}$ \citep{2005ApJS..160..149S}, which is higher than that of the most massive galaxies in Virgo -- M49 (981 km s$^{-1}$) and M87 (1284 km s$^{-1}$) \citep{2011MNRAS.413..813C}. As part of the Virgo southern extension, NGC 4532 and DDO 137 appear to be infalling into Virgo \citep{2014MNRAS.438.1922D}. At the TF distance, the projected separation between NGC 4532 and DDO 137 is {$\sim 48$ kpc}, which is double that of the local LMC/SMC system. However, unlike the Magellanic Clouds, there is no nearby massive parent galaxy -- only more distant galaxies and the massive Virgo cluster. Interactions between similar pairs of low-mass galaxies have been studied by \citet{2016MNRAS.459.1827P}, \citet{2022AJ....163...49L}, and \citet{2022ApJ...926L..15J}. The effect of the cluster environment on the properties of galaxy pairs in the WALLABY pilot survey is investigated in \citet{2023MNRAS.519..318K} and \citet{2024arXiv241022406H}.

Extended HI emission associated with the NGC 4532 / DDO 137 galaxy pair was first detected with the Arecibo telescope in 1989 \citep{1992ApJ...388L...5H,1993AJ....106...39H}. More detailed studies of the parent galaxies conducted with the VLA \citep{1999AJ....117..811H,2009AJ....138.1741C} suggested that 37\% of the HI lays outside of the visible galaxies. However, apart from three discrete clouds, the extended emission seen with Arecibo was resolved out. Deep CCD imaging in the $B$ and $R$ bands revealed no optical counterparts to the clouds.

The projected extent of the Arecibo cloud was measured to be 200 kpc, but later ALFALFA observations \citep{2008ApJ...682L..85K} revealed an even larger 500 kpc extent for the system, with a tail extending to the southwest of the galaxy pair {(these authors used the slightly larger distance of 16.5 Mpc)}. This feature is the longest and most massive HI tail structure so far found in the vicinity of the Virgo Cluster. The feature was described by \citet{2008ApJ...682L..85K} as being most consistent with a tidal origin, but the vicinity to the Virgo cluster, and analogy with the LMC/SMC system, suggests that the role of ram pressure should not be discounted \citep{2016MNRAS.459.1827P}.
Extended HI has been detected around many other galaxy systems, often serendipidously, and has been an excellent way to probe galaxy interactions in dense environments and  delivers clues as to the dominant mechanisms through which cold gas is stripped from galaxies, or accreted onto galaxies, and the subsequent impact on the star formation \citep{2021PASA...38...35C}. 

Examples of systems with extended HI in the form of tails include: Hickson Compact Group 44 \citep{2017MNRAS.464..957H} which possesses a 0.45 Mpc tail; at least 10 HI tails in the Virgo cluster 
\citep{ 2004MNRAS.349..922D, 2005A&A...437L..19O, 2007ApJ...659L.115C, 2007ApJ...665L..19H}; six tails discovered in the MeerKAT Fornax Deep Survey \citep{2023A&A...673A.146S}; tails associated with the NGC 4631/4656 system \citep{2023ApJ...944..102W}; tails and HI features associated with the M81 and Leo triplets \citep{1978AJ.....83..219R,1981MNRAS.195..327A,2018ApJ...865...26D,2022A&A...658A..25W}; a tail of gas and other extended features associated with NGC 3395/3396 \citep{1999MNRAS.308..364C,2024MNRAS.532.1744Y}; and two tails associated with galaxies near the A1367 cluster which are of similar size \citep{2012MNRAS.419L..19S,2022MNRAS.511..980S}. The origin of these tails is sometimes uncertain, but tidal interaction \citep{2005MNRAS.363L..21B,2008ApJ...673..787D}, ram pressure stripping \citep{2022ApJ...928..144L}, or a combination of the two is commonly invoked. See \citet{2022A&ARv..30....3B} for a review and further examples.

In this paper, we describe the new WALLABY observations in Section 2, present the results and a comparison with other data in Section 3, present a  dynamical model for the system in Section 4, including the relative role of hydrodynamical and tidal forces, and conclude in Section 5. For the purposes of parameterisation of length and mass scales, we assume a nominal NGC 4532 / DDO 137 distance of 13.8 Mpc, and a Virgo cluster distance of 16.5 Mpc.

\section{OBSERVATIONS}
\label{s:observations}

\begin{table*}
    \centering
    \begin{tabular}{lcccccc}
    \hline
    Parameter      & NGC 4532 & DDO 137 & Bridge & E1 & A1 & A2 \\
                   & \\
    \hline
    RA (J2000)$^{\rm a}$ & $12^{\rm h}34^{\rm m}19.3^{\rm s}$ & $12^{\rm h}34^{\rm m}45.3^{\rm s}$ & $12^{\rm h}34^{\rm m}29.6^{\rm s}$ & $12^{\rm h}34^{\rm m}48.5^{\rm s}$ & $12^{\rm h}34^{\rm m}04.2^{\rm s}$ & $12^{\rm h}34^{\rm m}11.5^{\rm s}$ \\
    DEC (J2000)$^{\rm a}$ &  $+06^{\circ}28'04''$ & $+06^{\circ}18'02''$ & $+06^{\circ}22'04''$ & $+06^{\circ}24'52''$ & $+06^{\circ}23'39''$ & $+06^{\circ}12'51''$ \\
    HI flux (Jy km s$^{-1}$)$^{\rm b}$ & $40.0\pm4.0$ & $6.5\pm0.6$ & $8.1\pm0.8$ & $1.1\pm0.1$ & $3.1\pm0.3$ & $4.2\pm0.4$ \\
    Velocity, $cz$ (km s$^{-1}$) & $2021\pm3$ & $2054\pm5$ & $1990\pm10$ & $1940\pm7$ & $1882\pm10$ & $1980\pm10$ \\
    Velocity width, $W_{50}$ (km s$^{-1}$) & $151\pm5$ & $52\pm7$ \\
    & \multicolumn{2}{c}{Kinematic fits} \\
    Velocity, $cz$ (km s$^{-1}$) & $2023\pm8$ & $2063\pm5$ \\
    Rotation velocity, $V_{\rm rot}$ (km s$^{-1}$) & $93\pm 10$ & $38\pm10$ \\
    Position angle & $347^{\circ}\pm4^{\circ}$ & $96^{\circ}\pm4^{\circ}$ \\
    Inclination & $74^{\circ}\pm4^{\circ}$ & $30^{\circ}\pm10^{\circ}$ \\
     & \multicolumn{2}{c}{Derived parameters} \\
     HI mass, $M_{\rm HI}$ (M$_{\odot}$)$^{\rm c}$ & $1.8\times10^9$ & $2.9\times10^8$ & $3.6\times10^8$ & $4.8\times10^7$ & $1.4\times10^8$ & $2.7\times10^8$ \\
    Total mass, $M_{\rm T}$ (M$_{\odot}$)$^{\rm c,d}$ & $1.5\times 10^{10}$ & $1.2\times 10^{9}$ \\
    Stellar mass, $M_*$ (M$_{\odot}$)$^{\rm c,e}$ & $1.5\times 10^{9}$ & $3.6\times 10^{8}$ \\
    Star formation rate, SFR (M$_{\odot}$ yr$^{-1}$)$^{\rm c,f}$ & 0.52 & 0.014 \\
    Star formation efficiency, ${\rm SFR}/M_{\rm HI}$ (yr$^{-1}$) & $2.9\times10^{-10}$ & $4.7\times10^{-11}$\\
    Depletion time, $M_{\rm HI}/{\rm SFR}$ (Gyr) & $3.4$ & $2.1$\\
    
    \hline
    \multicolumn{7}{l}{$^{\rm a}$ Positions for NGC 4532 and DDO 137 from NED. Positions for the HI clouds are indicative only, since they are diffuse features.}\\
    \multicolumn{7}{l}{$^{\rm b}$ Due to different measurement methodology, no phase 1 flux correction \citep{2022PASA...39...58W} was applied.}\\ 
    \multicolumn{7}{l}{$^{\rm c}$ Assuming a distance of 13.8 Mpc.}\\
    \multicolumn{7}{l}{$^{\rm d}$ Within the radius of the final rotation curve measurement in Figure~\ref{fig:rotcur}.}\\
    \multicolumn{7}{l}{$^{\rm e}$ From the SDSS extinction-corrected $z$-band magnitude and $g-z$ colour, using equation 6 of \citet{2020MNRAS.495..905R}.}\\
    \multicolumn{7}{l}{$^{\rm f}$ From the GALEX NUV \citep{2018ApJS..234...18B} and WISE W3
    \citep{2014AA...565A.128C} 
    flux densities, using equations 9 and 10 of \citet{2022MNRAS.510.1716R}.}\\
    \end{tabular}

    \caption{Measured and derived parameters for the composite system WALLABY J123424+062511, including the galaxies NGC 4532 and DDO 137, and four of the most prominent diffuse features in Figure~\ref{fig:moments}: the bridge which extends between the two galaxies; the eastern cloud E1 and the two arms A1 and A2, which extend south-west from each of the galaxies. Velocity widths and rotation velocities are measured in the rest frame, as is HI flux which is integrated over rest-frame velocity.}
    \label{params}
\end{table*}

A single 8.7-h observation was taken with ASKAP on 5 December 2019 of the northern part of the NGC 4636 field at the central coordinates RA 12:38:02.7, Dec +04:56:59 (J2000) and frequency range 1151.5 to 1439.5 MHz. The scheduling block numbers for the source and bandpass calibrator observations are 10736 and 10737, respectively. Only a single 36-beam {\tt square\_6x6} footprint \citep{Hotan2021}, with a beam spacing of $0.9^{\circ}$, was used in the observation. Normal WALLABY observations employ a second footprint at a slightly offset position to ensure a uniform sensitivity response across the field. Unfortunately, the data from the second observation on 7 December 2019 was not usable due to poor data quality. Therefore the data cube resulting from this observation has lower sensitivity and higher noise variation than a normal WALLABY data cube. 

The closest beams to NGC 4532 were beams 4 and 5, which had central pointings of RA 12:32:39, Dec +06:18:01 (J2000) and RA 12:36:15, Dec +06:18:00 (J2000), respectively. The separation of these beam centres from NGC 4532 is approximately 27 arcmin and 31 arcmin, respectively, which are both within the nominal ASKAP FWHP beam size of $1.09 \lambda/D$, or $1.1^{\circ}$ at 1.4 GHz. The sub-cube for the NGC 4532 / DDO 137 system was extracted from the fully-mosaiced cube of the field, which can also been downloaded from the CSIRO ASKAP Science Data 
Archive\footnote{\url{https://doi.org/10.25919/bp9s-0278}}, so also includes minor signal contributions from other beams. At the time of processing, a Gaussian approximation was used for primary beam correction. Compared with the later holographic beam estimates, this correction has since been shown to slightly overestimate flux densities, especially for positions distant from the beam centre and beams distant from the centre of the footprint.

\begin{figure*}
\begin{center}
\includegraphics[width=0.5\textwidth,angle=-0]{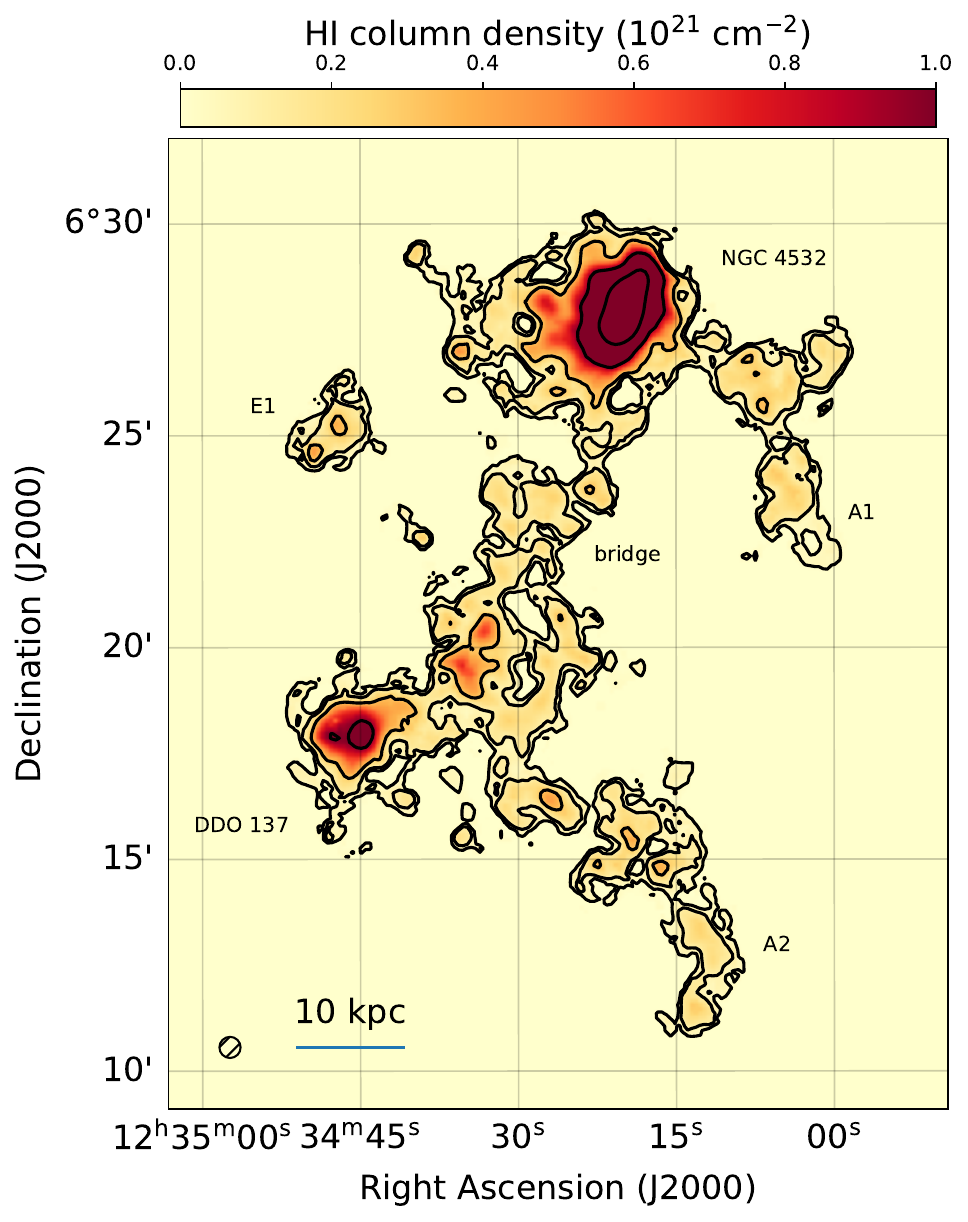}\includegraphics[width=0.506\textwidth,angle=-0]{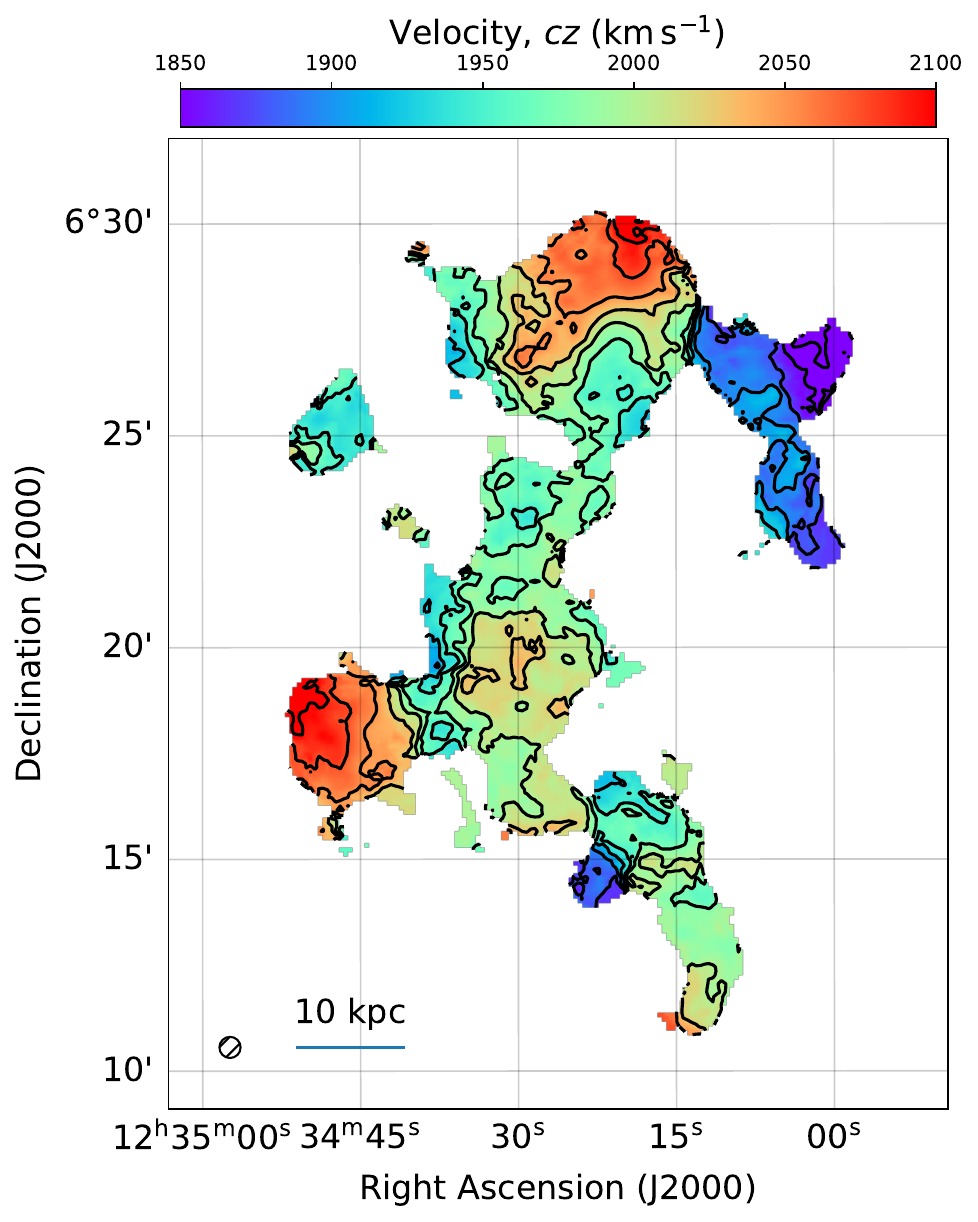}
\caption{{\it Left:} the WALLABY HI column density image of the NGC 4532/DDO 137 system (also known as WALLABY~J123424+062511).  The annotated features are discussed in the text. The contours levels are: 0.03, 0.1, 0.3, 1.0 and $3.0\times10^{21}$ cm$^{-2}$. The peak column density of $6.3\times10^{21}$ cm$^{-2}$ is at the position of NGC 4532.  {\it Right:} the mean HI velocity field with contours levels evenly spaced by 25~km s$^{-1}$ between 1850 and 2100~km s$^{-1}$.   Both images use masks generated by SoFiA \citep{2021MNRAS.506.3962W}. The 30 arcsec beam size {and a 10 kpc scale bar (assuming a distance of 13.8 Mpc)} are shown the lower left of each image.}
\label{fig:moments}
\end{center}
\end{figure*}

Only baselines shorter than 2 km are used for normal WALLABY observations, which results in an angular resolution of 30 arcsec after weighting and tapering. The cube from which the NGC 4532 / DDO 137 data was extracted has a frequency range of 1295.5 to 1439.5 MHz (frequencies lower than this were not processed owing to GNSS satellite interference), and a channel spacing of 18.519 kHz, or a velocity spacing of 3.9 km~s$^{-1}$ at the redshift of the system, and a pixel size of 6 arcsec. The rms at nearby positions and frequencies to NGC 4532 / DDO 137 is 2.6 mJy beam$^{-1}$ in each channel, which is around 1.6 times the nominal WALLABY sensitivity of 1.6 mJy beam$^{-1}$. Other specifications and WALLABY specific-details of the data processing pipeline can be found in \citet{2022PASA...39...58W}.

HI detections, moment maps, velocity fields and spectra for the NGC 4636 pilot phase 1 field were made using SoFiA \citep{2021MNRAS.506.3962W}. Due to residual continuum emission from the nearby strong radio source, 3C 273, SoFiA was not run in its normal `blind' mode, but was guided by a combined catalogue of known optical/HI sources within the footprint and redshift range. The combined NGC 4532 / DDO 137 system was merged into a single system by SoFiA, and catalogued as WALLABY~J123424+062511 \citep{2022PASA...39...58W}, where the coordinates refer to the HI centre-of-mass of the system, not to either of the optical centres. 

As the data in the vicinity of the NGC~4532 / DDO~137 system turned out to be largely unaffected by continuum residuals, we were able to carry out a much deeper, blind source finding run on a smaller $1^{\circ}\times 1^{\circ}$ region encompassing the entire system. By setting the detection threshold of SoFiA's standard {\tt S+C} source finding algorithm to a fairly low value of $3.4 \sigma$, we were able to pick up a lot of additional emission along the gas bridge and tidal arms of the NGC~4532
/ DDO~137 system. Hence, the analysis presented in this paper is based on our additional deep SoFiA run rather than the standard WALLABY data products.

{Measurements of column density and total flux for WALLABY observations (and resolved interferometric observations in general) are subject to at least two main uncertainties. The first is that the measured flux can be significantly in error if low signal-to-noise emission is not fully deconvolved. For most arrays, this often leads to an overestimate of flux density due to significant beam sidelobes when non-uniform {\it uv} weights are used. However for ASKAP, for which the 2-km core has an almost-Gaussian natural beam, the effect is less, and in the opposite sense. As discussed by \citet{2022PASA...39...58W} and \citet{2024PASA...41...88M}, the flux loss can be as much as $\sim 30$ per cent for fainter galaxies. The second uncertainty for extended low-column-density emission is that significant amount of flux density may be outside the mask, and therefore lost in the `noise'. This is discussed further in the next section, where comparisons are made with previous single-dish observations.}

\begin{figure}
\begin{center}
\includegraphics[width=1.1\columnwidth,angle=-0]{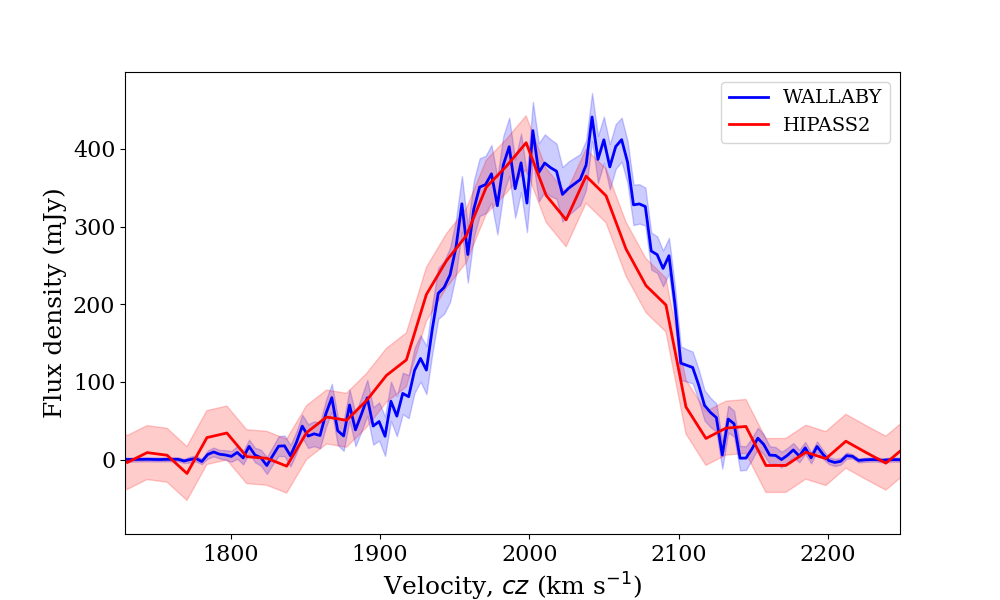}
\caption{The spatially integrated HI spectrum of WALLABY~J123424+062511 overlaid on a spatially integrated HIPASS spectrum. The shaded areas show the 1-$\sigma$ error bands.}
\label{fig:spectrum}
\end{center}
\end{figure}

\begin{figure*}
\begin{center}
\includegraphics[width=1.0\textwidth,angle=-0]{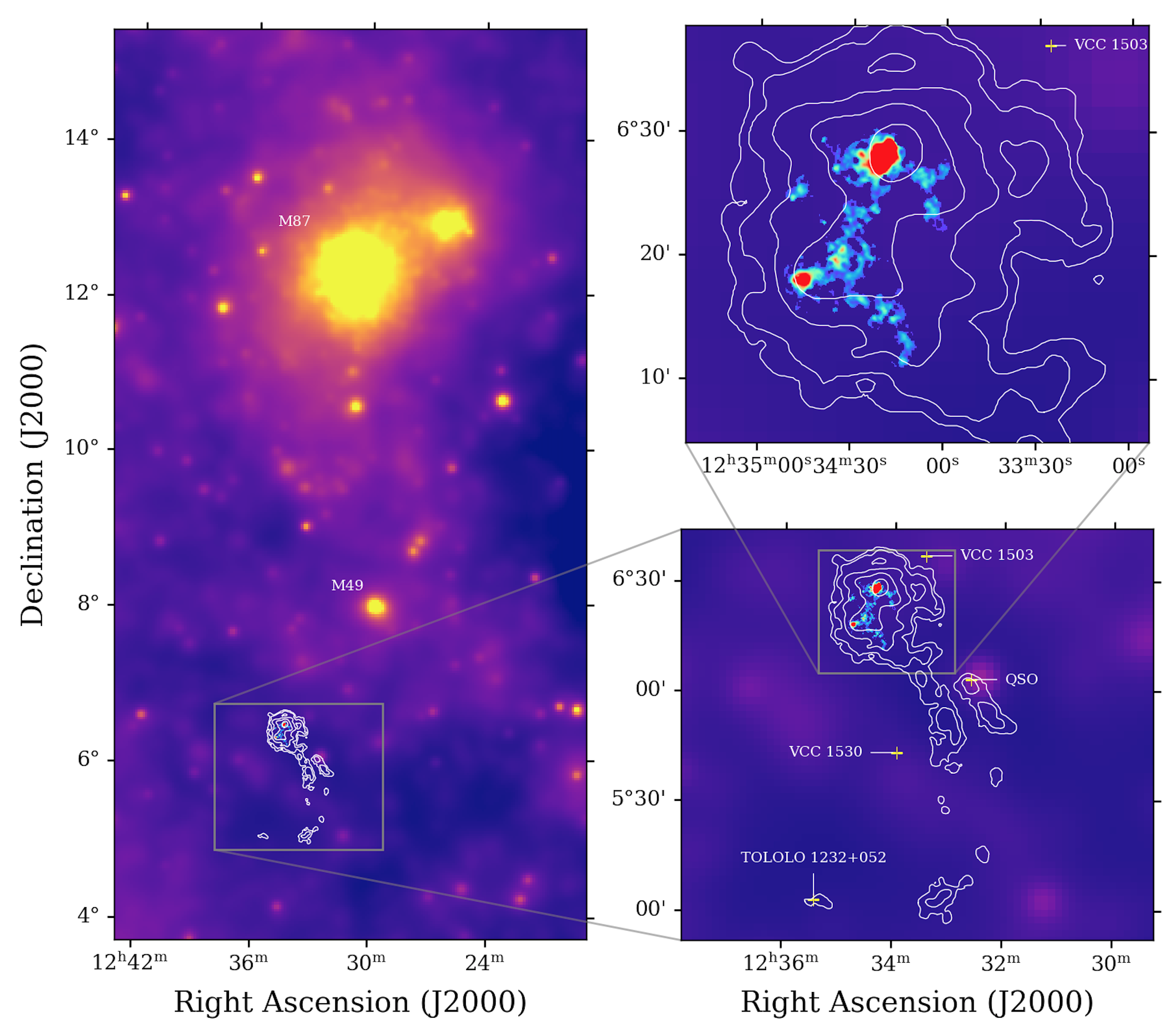}
\caption{{\it Left:} a ROSAT All-Sky Survey image of the Virgo cluster \citep{1994Natur.368..828B} overlaid with  ALFALFA HI contours of NGC 4532/DDO 137 \citep{2008ApJ...682L..85K}. The positions of the two massive elliptical galaxies M49 and M87 are marked. {\it Lower right:} an expanded image showing the ALFALFA HI contours for NGC 4532/DDO 137 and the HI tail overlaid on the WALLABY HI image. {\it Upper right:} a further expanded image showing the diffuse HI envelope around the WALLABY image of  NGC 4532/DDO 137. The contour levels from \citet{2008ApJ...682L..85K} are 0.96, 3.9, 15, 62 and $250 \times 10^{18}$ cm$^{-2}$. The peak WALLABY column density is $6.3\times10^{21}$ cm$^{-2}$. The positions of other galaxies with known optical velocities in the range 1700 to 2200 km s$^{-1}$ (VCC 1503, VCC 1530 and Tololo 1232+052) and the background QSO, SDSSJ123235.82+060310.0, are marked in the expanded images.
}
\label{fig:rosat}
\end{center}
\end{figure*}

\begin{figure}
\begin{center}
\includegraphics[width=1.0\columnwidth,angle=-0]{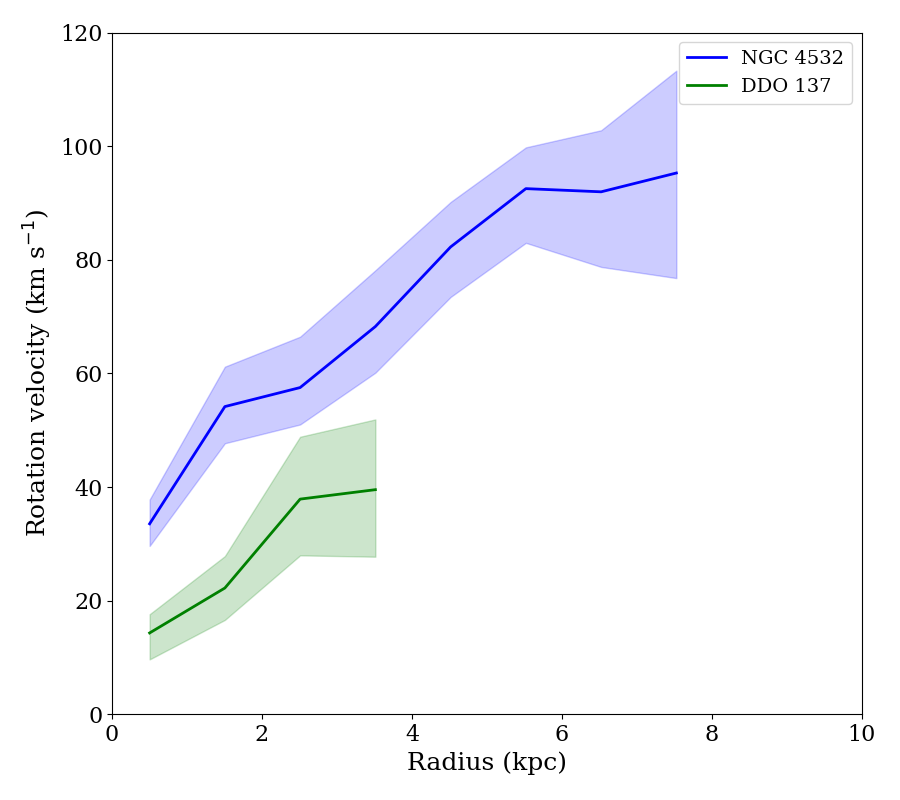}
\includegraphics[width=1.0\columnwidth,angle=-0]{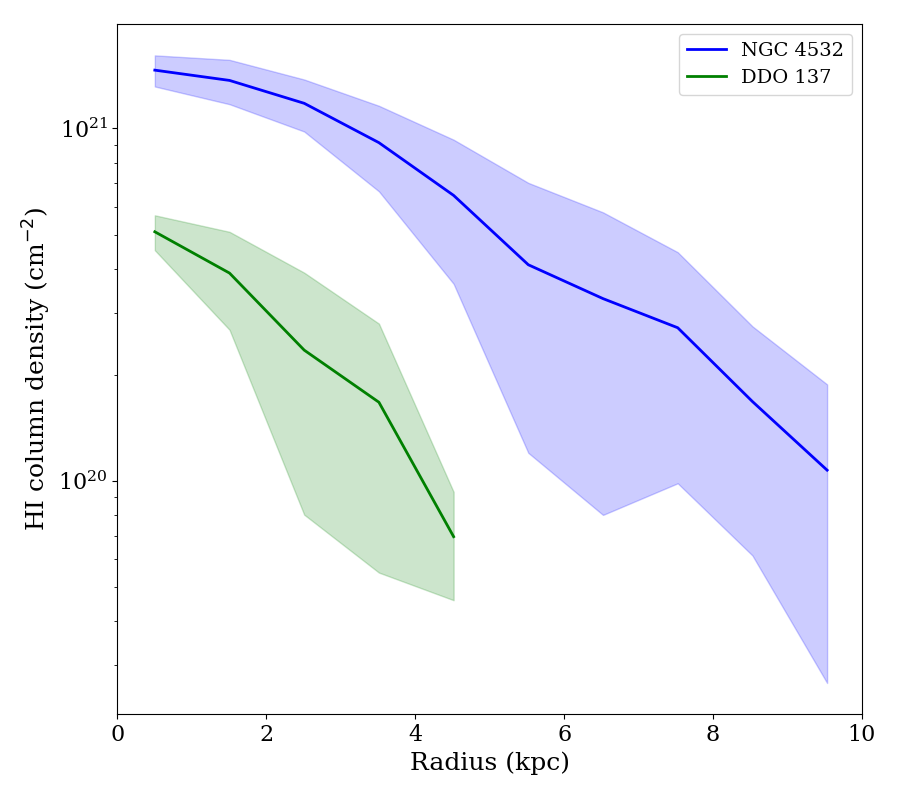}
\caption{Rotation velocity (top) and HI column density (bottom) for NGC 4532 (upper blue curves) and DDO 137 (lower green curves) from the kinematic fits, and elliptical ring fitting. The kinematic fit assumes disk-like rotation with constant inclination and position angle (parameters in Table~\ref{params}). The inclination-corrected surface density profile assumes that the HI is optically thin. The shaded areas show the 1-$\sigma$ error bands.
}
\label{fig:rotcur}
\end{center}
\end{figure}

\section{RESULTS}

Figure~\ref{fig:moments} shows the HI column density and velocity field for WALLABY J123424+062511. The two individual galaxies (NGC 4532 to the north and DDO 137 in the south east) are clearly detected. Both appear to have kinematic signatures of rotation with central (heliocentric) velocities of $cz=2021$ km~s$^{-1}$ and 2054 km~s$^{-1}$ for NGC 4532 and DDO 137, respectively (see Table~\ref{params}). The velocity difference of 33 km~s$^{-1}$ is similar to  the 18 km s$^{-1}$ difference quoted by \citet{2016MNRAS.459.1827P} and the 29 km s$^{-1}$ measured by \citet{2009AJ....138.1741C}. A bridge or filament of gas can also be seen to  extend without break between the two galaxies. The mean velocity of the bridge is approximately 1990 km~s$^{-1}$, lower than either of the individual galaxies, but closer to that of the more massive of the pair, NGC 4532. There are also three additional features that can be seen in Figure~\ref{fig:moments}, namely a small cloud (E1) in the east and two arm-like features (A1 and A2) which are breaking away to the south-west from NGC 4532 and the bridge, near the position of DDO 137. The earlier VLA observations of \citet{1999AJ....117..811H} \citep[also presented by][]{2009AJ....138.1741C} detect E1 and part of A1 (A and C in their nomenclature, respectively) as well as part of the bridge near DDO 137.  

The basic parameters of the galaxies, bridge, cloud E1 and arms A1 and A2, as measured from the WALLABY data, are given in Table~\ref{params}. As already mentioned, the velocities for the galaxies are consistent with previous observations with high accuracy. The HI flux for NGC 4532 of 40.0 Jy km~s$^{-1}$ lies between the VLA measurements quoted by \citet{1999AJ....117..811H} and \citet{2009AJ....138.1741C} of $32.4\pm2.3$ and 50 Jy km~s$^{-1}$, respectively. Similarly, the WALLABY flux for DDO 137 of 6.5 Jy km~s$^{-1}$ lies between the VLA measurements by the same authors of $5.2\pm0.5$ and 10.1 Jy km~s$^{-1}$, respectively. The differences probably arise from how much of the extended HI envelopes for the two galaxies are included in the measurements. The ALFALFA pipeline measurements, which include all flux within an isodensity ellipse fitted at 50\% of the peak column density, are 41.56 and 11.97 Jy km~s$^{-1}$ for NGC 4532 and DDO 137, respectively \citep{2018ApJ...861...49H}.

Figure~\ref{fig:moments} shows that the extremal mean velocities in the system lie in arm A1 to the west of NGC 4532 (1880 km~s$^{-1}$), the north edge of NGC 4532 (2105 km~s$^{-1}$), and the eastern edge of DDO 137 (also 2105 km~s$^{-1}$). The integrated spectrum in Figure~\ref{fig:spectrum} shows that, due to velocity dispersion about the mean local velocities, the whole system extends across a wider range in velocity, from 1840 to 2160 km~s$^{-1}$, with the FWHM extent of the integrated spectrum being 153 km~s$^{-1}$. 
{Figure~\ref{fig:spectrum} also compares the WALLABY spectrum with a HIPASS spectrum (re-reduced to mitigate scanning sidelobes from NGC 4535), spatially integrated over a 44 arcmin region. The flux densities and velocity-integrated fluxes are in good agreement (the latter is $63.6\pm1.0$ Jy km~s$^{-1}$ for the WALLABY spectrum, compared with $62.1\pm2.9$ Jy km~s$^{-1}$ from HIPASS). There is a minor deficit in low-velocity gas at $\sim 1920$ km~s$^{-1}$ in the WALLABY spectrum and a deficit of high-velocity gas at $\sim 2070$ km~s$^{-1}$.  
The former is probably due to diffuse low-column density gas to the east and west of the bridge which, from the ALFALFA velocity field \citep{2008ApJ...682L..85K}, has a lower velocity than either of the two galaxies, or the bridge itself. The deficit of high-velocity gas \citep[also seen in the original HIPASS spectrum;][]{2006MNRAS.371.1855W} may be due to residual scanning sidelobes}.
The single-dish observation of \citet{1981ApJS...47..139F} indicates an HI flux which is similar to that of HIPASS and WALLABY, $64.4\pm4.6$ Jy km s$^{-1}$ within the 22 arcmin beam of the NRAO 43-m telescope, centred on NGC 4532. 

The WALLABY image is overlaid with the Arecibo ALFALFA HI contours in Figure~\ref{fig:rosat} , which also shows both in the context of an X-ray image of the Virgo cluster from the ROSAT All Sky Survey \citep{1994Natur.368..828B}. As noted by \citet{2008ApJ...682L..85K}, there is a positional and redshift association of their cloud 8 (the south-eastern cloud in Figure~\ref{fig:rosat}) with compact blue galaxy, Tololo 1232+052, which is too faint to be seen with WALLABY. If this cloud is part of the tail, as it seems to be from the velocity association with the other southern clouds, then Tol 1232+05 has either formed from the baryonic collapse of cool gas in the tail, or has benefited from the accretion of fresh gas from the tail. Although the former are commonly known as `tidal dwarf galaxies' \citep[TDGs; e.g.][]{1992Natur.360..715B}, we demonstrate in Section~\ref{s:discussion} that a mixed tidal/ram pressure origin for gas in the tail is more plausible. The only other nearby galaxies with known velocities in the range 1700 to 2200 km s$^{-1}$ are the nucleated dwarf ellipticals, VCC 1503 and 1530 \citep{1985AJ.....90.1681B} which have SDSS velocities in the range 1704 to 1749 km s$^{-1}$ \citep{2009ApJS..182..543A}.

\begin{figure}
\begin{center}
\includegraphics[width=\columnwidth,angle=-0]{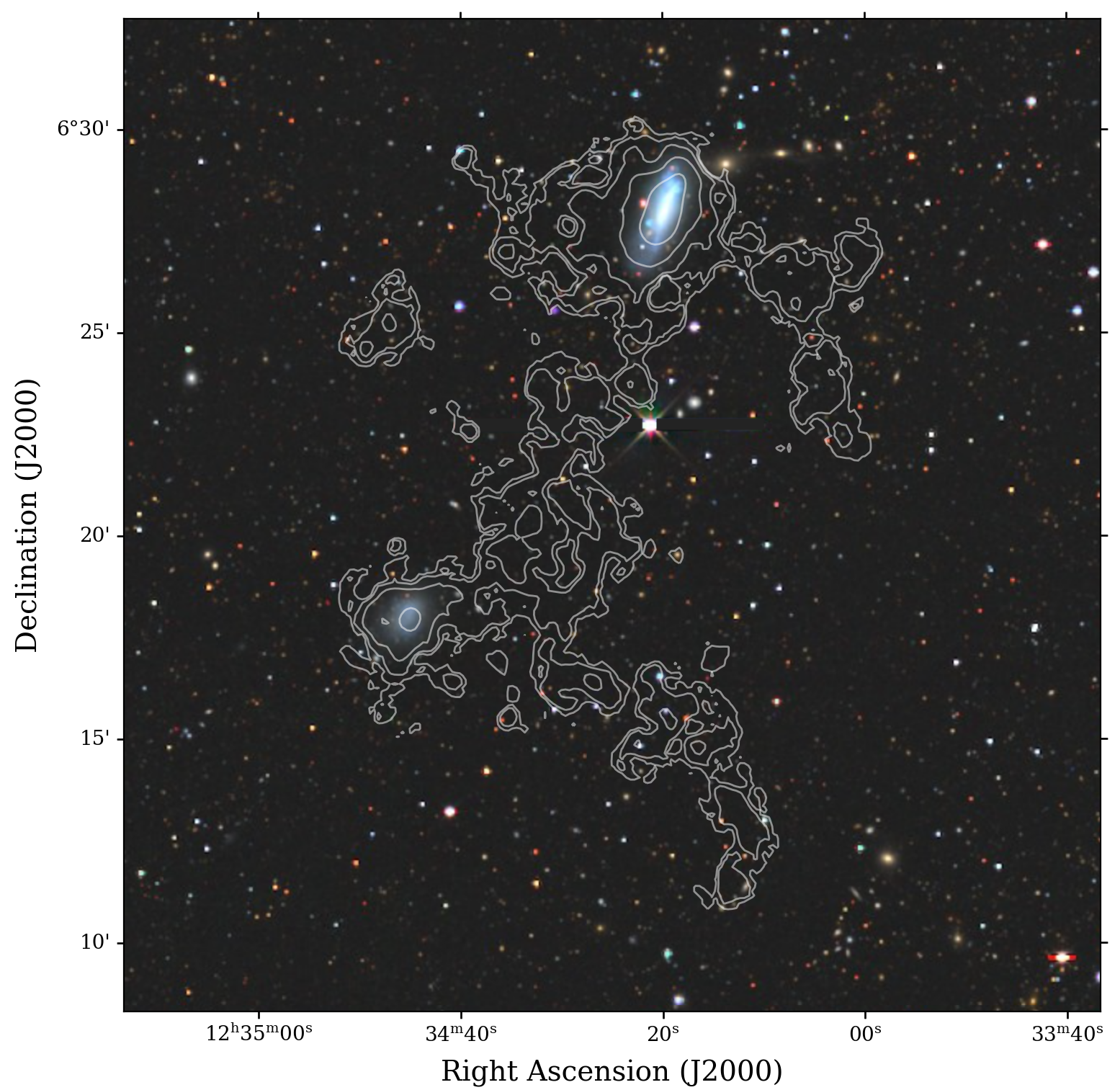}
\caption{Contours of the WALLABY HI column density for the NGC 4532/DDO 137 system overlaid on a DESI Legacy DR9 {\it grz} cutout \citep[][composite credit: Legacy Surveys/D. Lang, Perimeter Institute]{2019AJ....157..168D}. The contour levels are 0.3, 1, 3, 10 and $30\times 10^{20}$ cm$^{-2}$.
}
\label{fig:overlays}
\end{center}
\end{figure}

\begin{figure*}
\begin{center}
\includegraphics[height=0.5\textwidth,angle=-0]{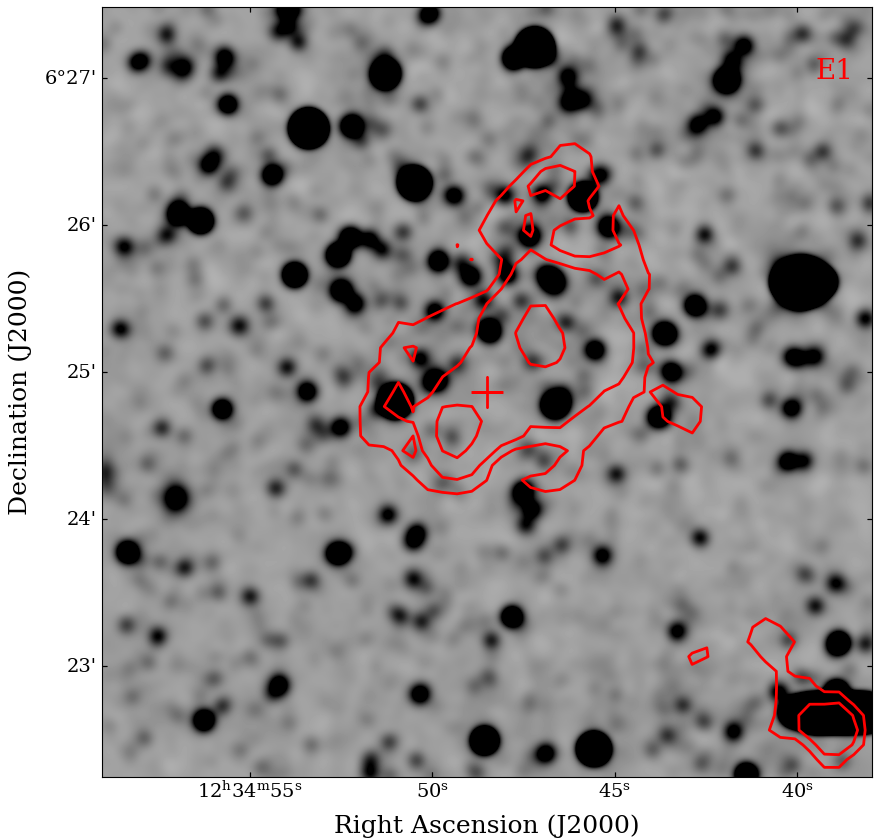}
\includegraphics[height=0.5\textwidth,angle=-0]{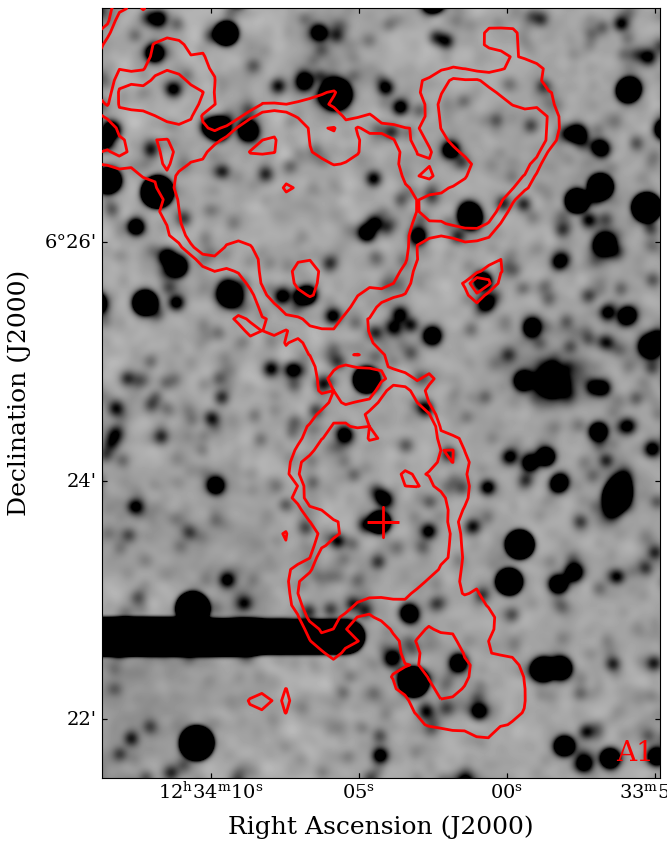}
\includegraphics[height=0.5\textwidth,angle=-0]{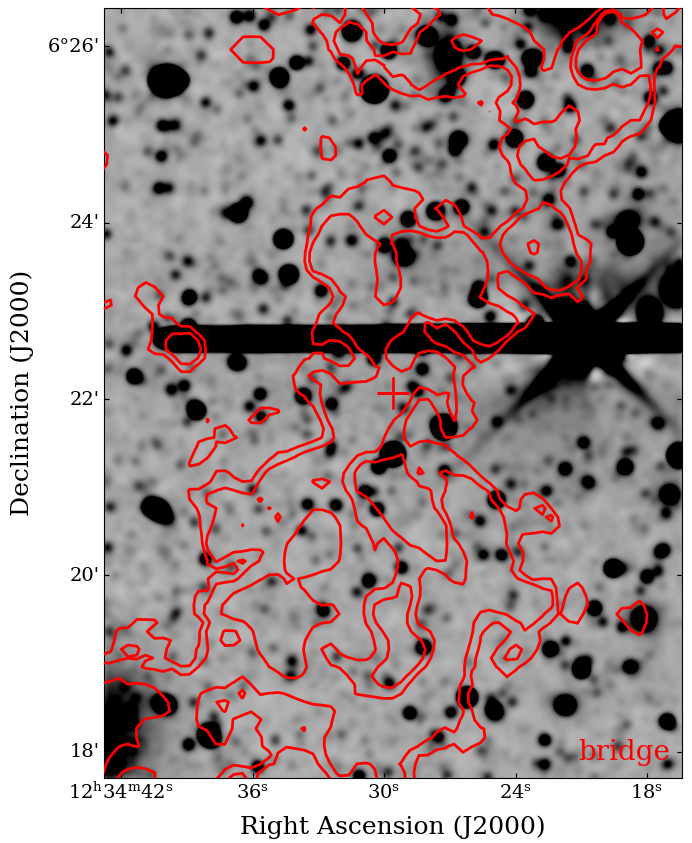}
\includegraphics[height=0.5\textwidth,angle=-0]{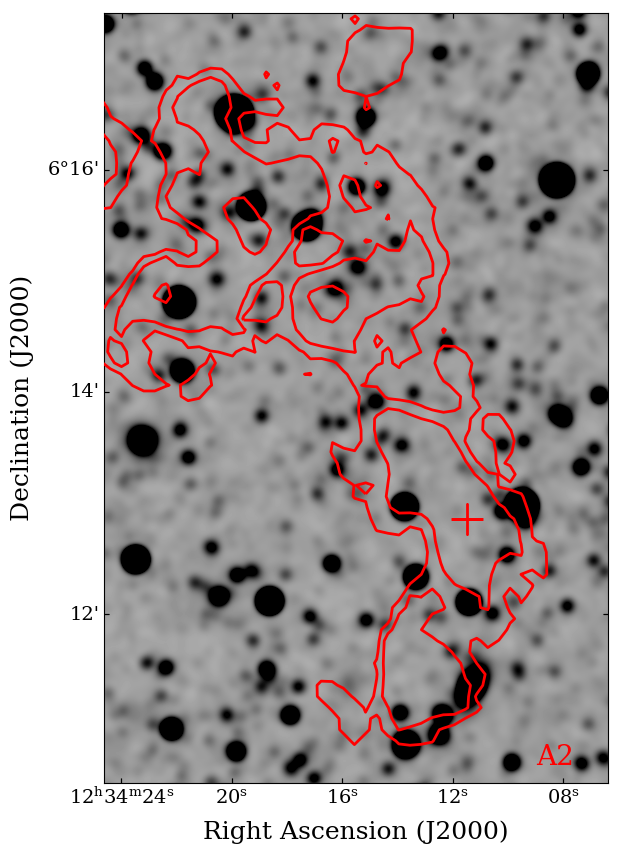}
\caption{{Contours of WALLABY HI column density for four extended components of WALLABY J123424+062511 (from top-left to bottom-right: cloud E1, arm A1, bridge and arm A2) overlaid on stacked DESI Legacy DR10 {\it griz} cutouts \citep{2019AJ....157..168D}. The DESI images have been convolved with a Gaussian of FWHM 6.2 arcsec. The $3\sigma$ surface brightness sensitivity is 27.7 mag arcsec$^{-2}$. The contour levels are the same as in Figure~\ref{fig:overlays}. The nominal coordinates of each component from Table~\ref{params} is marked with a red cross.}
}
\label{fig:zooms}
\end{center}
\end{figure*}

Clouds E1, A1 and A2 in Figure~\ref{fig:moments} are embedded in the larger halo of gas mapped with the Arecibo telescope by \citet{1992ApJ...388L...5H} and \citet{2008ApJ...682L..85K}, but which is too faint for WALLABY to detect -- see Figure~\ref{fig:rosat}. \citet{1992ApJ...388L...5H} mapped an $18'\times26'$ region around the galaxy pair and quote a galaxy+envelope flux of 183 Jy km s$^{-1}$, later updated to 97.4 Jy km s$^{-1}$ \citep{1999AJ....117..811H}. 
The ALFALFA study of \citet{2008ApJ...682L..85K} quotes 85.8 Jy km~s$^{-1}$, plus a further 5.9 Jy km s$^{-1}$ contained in the tail. \citet{2019AJ....157..194H} quote 71.28 Jy km~s$^{-1}$ from the same data with a further 8.37 Jy km~s$^{-1}$ contained in {\it ``the extension to the southwest''}, which contains the bulk of the HI mass in the tail (although they also confusingly quote a much higher individual flux for NGC 4532 than the combined system).  
Depending on which ALFALFA measurement is used, the WALLABY data therefore seems to miss around 12-35\% of the HI in the low density envelope around the system, which is easily explained by lack of sensitivity to low column density diffuse gas. However, it is puzzling that other single-dish observations have not picked up much of this component. Some of this could be due to the low column density nature of the HI envelope. It is also possible that the Arecibo fluxes are slightly over-estimated and/or the errors are underestimated due to the significant amplitude of sidelobes that were present on this telescope. \citet{1983AJ.....88..272H} suggest an accuracy of 25\% for pre-ALFA HI fluxes measured for resolved galaxies. \citet{2011ApJS..194...20P} discuss the sidelobe efficiency of the ALFA receiver, which is 20\% for the first sidelobe alone. 

Although it remains for future observations with ASKAP, FAST or MeerKAT to provide more accurate measurements of the diffuse low-column density gas, it is nevertheless worth noting that the sum total of all HI mass in the diffuse features outside the main bodies of NGC 4532 and DDO 137 (bridge, arms, tail etc.) is $\sim 2\times 10^9$ M$_{\odot}$, which is about the same as the HI mass contained in the individual galaxies ($2.1\times 10^9$ M$_{\odot}$, from Table~\ref{params}). With gas fractions of $f_{HI} = 1.2$ and $0.8$ for NGC 4532 and DDO 137, respectively, both galaxies currently lie within the gas-fraction main sequence (or its extrapolation in the case of DDO 137) of xGASS star-forming galaxies defined by \citet{2020MNRAS.493.1982J}. If the diffuse gas were to be added back to each galaxy in equal measure, the resultant HI gas fractions would then be $\sim 0.4$ dex higher than the same median main sequence, though still well within the scatter of WALLABY infall and field galaxies near the Hydra Cluster \citep{2022MNRAS.510.1716R}, which is better matched in HI and stellar mass range than xGASS \citep{2018MNRAS.476..875C}.

Kinematic fits are not directly available from the WALLABY pilot survey phase 1 data release \citep{2022PASA...39...58W, 2022PASA...39...59D} as WALLABY J123424+062511 is classified as a complex system. Nevertheless, we have been able to extract the individual velocity fields for the two galaxies and fit them using
3D-Barolo \citep{2015MNRAS.451.3021D}. For this purpose, a 5-$\sigma$ mask was created to eliminate the extended features in the WALLABY data cube. This corresponds to an HI column density of $6\times 10^{19}$ cm$^{-2}$ per 3.9 km s$^{-1}$ channel. The kinematic centroid was fixed at the optical coordinates listed in Table~\ref{params}. Following the earlier kinematic analysis of \citet{1999AJ....117..811H}, and the available axis ratio measurements listed in NED, initial values for the inclination of $72^{\circ}$ and $30^{\circ}$ were used for NGC 4532 and DDO 137, respectively. The final fit values from 3D-Barolo are $74^{\circ}\pm4^{\circ}$ and $30^{\circ}\pm10^{\circ}$. Within the resolution of the WALLABY data, and within the tight mask around the galaxies, there was only a small deviation at different radii from these values, so a flat disk model was adopted for both galaxies. The inclination for DDO 137 is unfortunately barely constrained by the data, as is often the case for low-mass galaxies with limited flattening of their outer rotation curves. The other kinematic parameters are listed in Table~\ref{params} and the rotation curves and surface density profiles are shown in Figure~\ref{fig:rotcur}. The surface density profiles are approximately exponential but, due to the mask, have been truncated below 0.7-$1.0\times 10^{20}$ cm$^{-2}$. The `flat' part of the rotation curves are at $93\pm10$ km~s$^{-1}$ and $38\pm10$ km~s$^{-1}$, for NGC 4532 and DDO 137 respectively. Within the last measured point, the total mass ratio is $M_{\rm T}$(NGC 4532)/$M_{\rm T}$(DDO 137) = $13^{+11}_{-5}$, with the total mass of NGC 4532 being $1.5\times 10^{10}$ M$_{\odot}$ within a radius of 7.5 kpc. Other derived parameters (HI mass and stellar mass) are also listed in Table~\ref{params}. In many respects, the properties of NGC 4532 and DDO 137 presented in Table~\ref{params} mirror those of the LMC and SMC \citep[see][]{2016ARA&A..54..363D}, a point to which we discuss in the following Section where we consider the origin of the bridge, clouds, arms and tail.

\section{DISCUSSION}
\label{s:discussion}

{The WALLABY column density contours are overlaid on a DESI Legacy DR9 $grz$ cutout in Figure~\ref{fig:overlays}}. The column density contours peak at the position of NGC 4532 and DDO 137, and their velocity fields are indicative of disturbed rotation. {Expanded versions of the extended HI features identified in Figure~\ref{fig:moments} (i.e.\ cloud E1, arms A1 and A2, and the bridge) are presented in Figure~\ref{fig:zooms} overlaid on smoothed, high-contrast stacked DESI $griz$ images. The measured $griz$ $3\sigma$ sensitivities of the smoothed images are 28.6, 27.6, 27.1 and 27.1 mag arcsec$^{-2}$, respectively, and the $3\sigma$ sensitivity of the unweighted and smoothed $griz$ stack is 27.7 mag arcsec$^{-2}$.
Despite this, no plausible optical/IR counterparts are seen}. Many small and faint galaxies are visible, but their spectroscopic and photometric redshifts place all of them firmly in the background. The HI and optical data are significantly deeper than presented by \citet{1999AJ....117..811H}, but the conclusions are similar.

The question of the origin of the diffuse envelope around NGC 4532/DDO 137 has been discussed by previous authors. 
\citet{2016MNRAS.459.1827P} appear to favour a ram pressure origin with NGC 4570 as the host galaxy, being the nearest  
massive ($M_* > 10^{10}$ M$_{\odot}$) neighbour. They compare NGC 4532/DDO 137 with the Magellanic System, but note the lack of ionized gas in HST QSO absorption-line observations compared with the Magellanic Stream. \citet{2008ApJ...682L..85K} also discuss ram-pressure interaction, but find that the expected amplitude of the ram pressure due to the intercluster medium (ICM) is 2--25 times smaller than the restoring force of the interstellar medium (ISM) in the galaxies. 

\citet{1992ApJ...388L...5H,1999AJ....117..811H} discuss tidal interactions between the pair. In such a scenario, the probable fate of the gas (and the galaxies themselves) is a merger into a single system. \citet{2008ApJ...682L..85K} also favour tidal interaction but, in order to explain their discovery of the 0.5 Mpc tail associated with the system, invoke a high-speed tidal interaction. Such a high-speed (1100 km s$^{-1}$) tidal interaction was also invoked by \citet{2008ApJ...673..787D} to explain the 0.25 Mpc HI tail extending from NGC 4254 in the Virgo cluster \citep[see also][]{2005A&A...439..921V}. High pairwise velocities will be common within and slightly beyond the virial radius of the Virgo cluster.

In the remainder of this section, we separately discuss the merits of quantitative tidal and ram pressure models and suggest that both are required to match the observational data.

\begin{table}
    \centering
    \begin{tabular}{lc}
    \hline
    Parameter      & Value  \\
                   & \\
    \hline
    \multicolumn{2}{c}{Initial parameters} \\
    Position$^{\rm a}$ (Mpc) & (1.19, 0.76, 0)   \\
    Velocity$^{\rm a}$ (km s$^{-1}$) & $(258, 0, 0)$  \\
    Stellar mass (NGC 4532) (M$_{\odot}$)  & $9 \times 10^{9}$ \\
    HI mass (NGC 4532) (M$_{\odot}$) & $3.6 \times 10^{9}$ \\
    Stellar mass (DDO 137) (M$_{\odot}$)  & $1.8 \times 10^{9}$ \\
    Halo mass (M$_{\odot}$) &  $3 \times 10^{11}$ \\
    Virial radius (kpc) & 134  \\
    Concentration parameter & 12 \\
    Binary separation (kpc) & 77 \\
    Eccentricity & 0.8 \\
    Virgo mass (M$_{\odot}$) & $10^{14}$  \\
     & \\
    \multicolumn{2}{c}{Final parameters} \\
    Position$^{\rm a}$ (Mpc) & $(-0.45, -1.88, 0)$  \\
    Velocity$^{\rm a}$ (km s$^{-1}$) & $(86, -211, 0)$  \\
    HI mass (M$_{\odot}$) & $2.8 \times 10^{9}$ \\
    Stellar mass (M$_{\odot}$) & $1.8 \times 10^{10}$ \\
    HI mass (tail) (M$_{\odot}$) & $2.5 \times 10^{8}$ \\
    New stellar mass (tail) (M$_{\odot}$) & $2.8 \times 10^{5}$ \\
    \hline
    \multicolumn{2}{l}{$^{\rm a}$Cartesian $(X,Y,Z)$ Virgocentric frame} \\
    \end{tabular}
    \caption{Initial and final (after 6 Gyr) parameters for an SPH model with no Virgo cluster medium which reproduces some of the observed parameters of WALLABY J123424+062511, but requires an unlikely passage through the Virgo cluster to generate a sufficiently long tail.
    }
    \label{tideparams}
\end{table}

\begin{figure*}
\begin{center}
\includegraphics[width=0.5\textwidth,angle=-0]{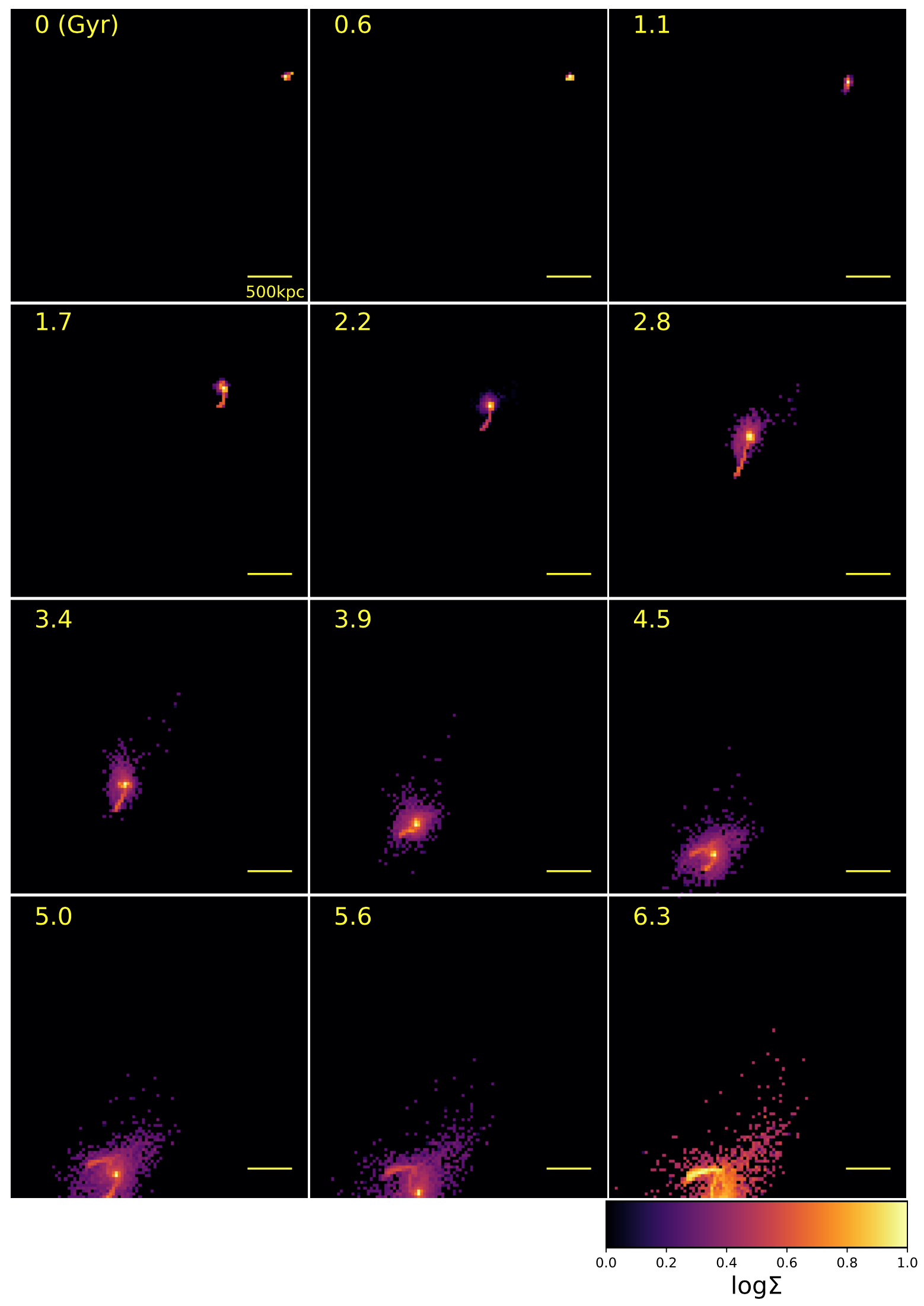}\includegraphics[width=0.5\textwidth,angle=-0]{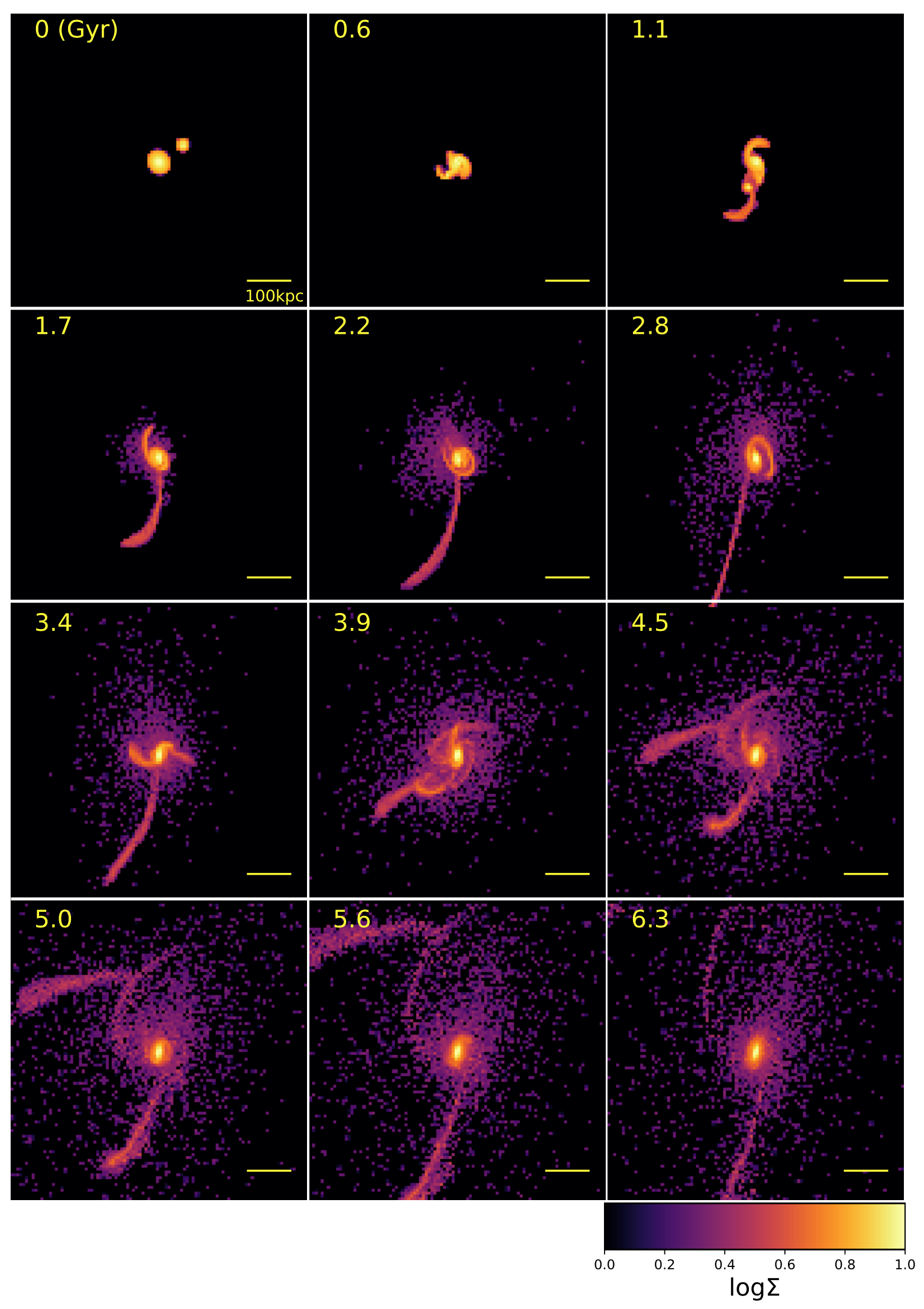}
\caption{{The distribution of gas particles in a SPH simulation (in arbitrary units of surface density) at 12 snapshots during a close passage of NGC 4532/DDO 137 past the Virgo cluster. The sub-plots are normalised in amplitude to unity and plotted with the logarithmic transfer function shown in the lower colour bars. The left hand $3\times 4$ sub-panels show a 3.3 Mpc box centred on the Virgo cluster; the right-hand $3\times 4$ sub-panels show a $5\times$ expanded box of width 0.66 Mpc centred on NGC 4532. The sub-panel labels are the time since the start of the simulation in Gyr. No Virgo cluster medium and no large-scale ram pressure forces are included in this simulation.}
}
\label{fig:sim}
\end{center}
\end{figure*}

\subsection{Tidal model}
\label{sec:tidal}

{It is relevant here to explore three aspects of tidal models. The first, as noted above, is to explore the mutual tidal interaction of NGC 4532 and DDO 137 which may be responsible for distributing gas around the WALLABY J123424+062511 system. The second is to explore the possibility that tidal fields (from more distant sources) can contribute to the non-merging of the two galaxies. The final aspect is to examine the role, if any, of tidal interactions in explaining the formation of the long tail.}

Regarding the origin of tidal fields, it is instructive to note that the tidal field of the Virgo cluster is at least similar in strength to that of NGC 4570, considered by \citet{2016MNRAS.459.1827P} to be the candidate host galaxy. Using abundance matching, these authors assume that NGC 4570 has a halo mass of $2.5\times 10^{12}$ M$_{\odot}$. However, its rotation velocity is only 150 km s$^{-1}$ \citep{1998MNRAS.298..267V} which is much less than the Milky Way, so a halo mass of $10^{12}$ M$_{\odot}$, or below, seems more likely. 
In this scenario, the cube of the ratio of the projected distances from NGC 4532 to Virgo and to NGC 4570\footnote{The NED velocity for NGC 4570 is lower by 236 km s$^{-1}$, and the mean of measured distances is higher by 2.7 Mpc.} is $(1.5/0.29)^3\approx140$, which is comparable to the Virgo/NGC 4570 mass ratio of $1$ -- $4\times 10^{14}/10^{12} = 100$ -- $400$, depending on whether we use
the $M_{200}$ mass for Virgo \citep{2017MNRAS.469.1476S} or the sum of the A and C components \citep{2018A&A...614A..56B} (the $\beta$-model discussed in Section~\ref{sec:ram} gives an intermediate estimate of the enclosed mass of $3.1\times 10^{14}$ M$_{\odot}$). However, given that the projected length of the tail is much larger than the distance to NGC 4570, the more likely origin for a distant tidal field is Virgo.

To investigate a tidal interaction model we utilised the smoothed-particle hydrodynamic (SPH) code as described in \citet{2013MNRAS.432.2298B} to reproduce the observed system. An array of initial parameters was explored using low-resolution simulations consisting of star and gas particles of initial mass $9\times 10^4$ M$_{\odot}$ and $3.6\times 10^4$ M$_{\odot}$, respectively. These simulations were conducted with and without the gravitational influence of the Virgo cluster. The simulations were not cosmological in nature, in that we assumed no ongoing mass accretion onto the Virgo cluster, which was modelled using a smooth and fixed potential. No Virgo cluster medium was included.
The parameters which were investigated were: initial position and velocity, the HI mass, stellar mass and dark halo mass (and size). 

Firstly, without the influence of the Virgo cluster, it was possible to produce long tidal arms, but only by extending the simulation to 6.3 Gyr. However, in all cases, NGC 4532 and DDO 137 had already merged by the end of the simulation. An external tidal field is required to prevent merging. This will result in further expulsion of loosely-bound gas from the system. 

{For simulations with the presence of Virgo, both leading and trailing arms are produced. These structures begin to form at 1.1 Gyr.
For the simulated orbit relative to Virgo which resulted in the closest match to the current observational parameters (see Table~\ref{tideparams}), we ran a high-resolution simulation with a mass resolution of $1.8 \times 10^{4}$ M$_{\odot}$ and a size resolution of 86 pc. This shown in the left-hand panel of Figure~\ref{fig:sim}. A $5\times$ expanded view is shown in the right hand panel in the same coordinate system, but centred on NGC 4532.}

{As seen in Figure~\ref{fig:sim}, the leading arm grows more quickly as the system approaches Virgo, but after the Virgo pericentre at 3 Gyr, both arms retract as the whole system puffs up. Once past Virgo, both arms re-grow to similar length of around 0.5 Mpc at 6.2 Gyr. Due to the configuration of the system, the leading and trailing arms are not parallel with each other.}

{However, this model is not tenable either. Its major drawback is that passage through the Virgo cluster is required due to the multi-Gyr formation timescale of the tail. Survival of a neutral HI tail would not be possible in the harsh cluster medium for this length of time. Observations and simulations have consistently shown that gas removal in clusters can be rapid, and that galaxies that fall into clusters lose much of their extended gas within a single radial round-trip \citep{2014PNAS..111.7914C,2018ApJ...865..156J}. Nevertheless, an important takeaway of the tidal model is that binary merging is suppressed, so that feeding of the less-extended circumgalactic gaseous halo can be maintained for a longer time.}

\begin{figure}
\begin{center}
\includegraphics[width=1.0\columnwidth,angle=-0]{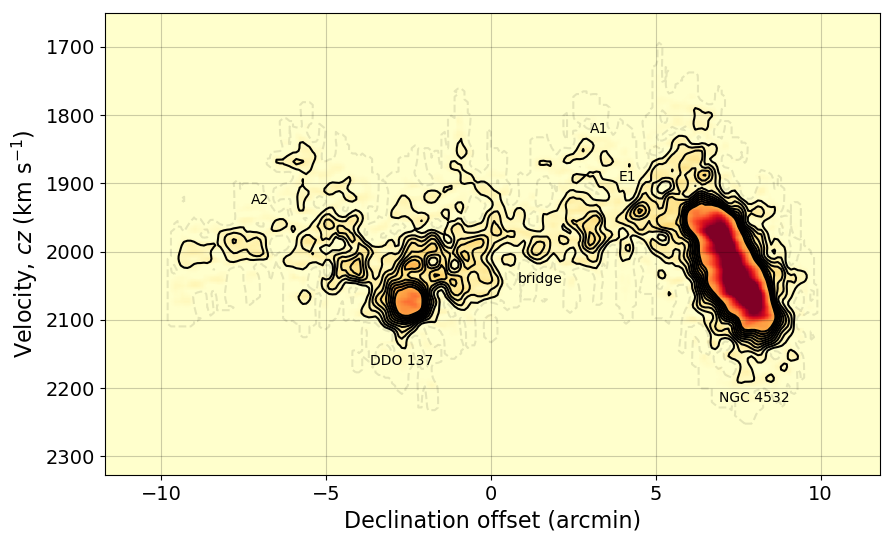}
\caption{{A position-velocity (PV) map of the HI emission in the NGC 4532/DDO 137 system. The horizontal axis is the Dec offset in arcmin from Dec (2000) = $6.3428^{\circ}$ and the vertical axis is the barycentric recession velocity, $cz$ in km s$^{-1}$. The HI emission has been spatially integrated along the whole RA axis in the masked cube. The PV image has been convolved with a Gaussian of standard deviation 8 km s$^{-1}$ to enhance fainter emission. The features are identified using the same nomenclature as in Figure~\ref{fig:moments}.}
}
\label{fig:PV}
\end{center}
\end{figure}

\subsection{Ram Pressure}
\label{sec:ram}

To better explain the large-scale HI features in the system, we therefore explore the addition of ram pressure from an  extended ICM. Since Virgo is a young, unrelaxed X-ray-bright cluster \citep{1994Natur.368..828B} with a shallower density profile than most evolved clusters, it seems appropriate to explore this cluster, rather than NGC 4570 \citep[as proposed by][]{2016MNRAS.459.1827P}, as the source of ram pressure. \citet{2024A&A...689A.113M} use data from the first five eROSITA surveys to estimate a cluster density profile $n_{\rm e}\approx 5\times10^{-5} (r/{\rm Mpc})^{-1.25}$ cm$^{-3}$.  \citet{2011MNRAS.414.2101U} use XMM-Newton spectroscopy along a northerly direction (away from WALLABY J123424+062511) to deduce an almost identical profile. This would imply a mean density at the projected distance of 1.5 Mpc for WALLABY J123424+062511 of up to $n_e\approx 3\times10^{-5}$ cm$^{-3}$. The $\beta$-model of \citet{2001ApJ...561..708V} used by \citet{2017ApJ...838...81Y} predicts a similar density. Extrapolation of the southern Suzaku profile \citep{2017MNRAS.469.1476S} suggests a slightly lower density ($\sim 10^{-5}$ cm$^{-3}$). 
\citet{2024A&A...689A.113M} find that, once galaxies are removed, the ratio of mean to median azimuthal X-ray flux (the emissivity bias) is $\sim 1.2$ in the cluster outskirts. Therefore, gas clumping will have a minor effect on estimated densities. However, as discussed below, the effect of geometry (the true separation of WALLABY J123424+062511 from the cluster centre) will mean that the actual ICM density will likely be lower by a factor of $\sim 2.5$.
Nevertheless, densities of this order represent the threshold at which rapid quenching, with a timescale of $\leq 1$ Gyr is seen to ensue in low-mass clusters in the C-EAGLE simulations of \citet{2022MNRAS.511.3210P}. An additional factor in favour of ram pressure is the large velocity difference ($737$ km~s$^{-1}$) between NGC 4532 and M87. Although high, this is not inconsistent with simulations of $z=0$ low to intermediate-mass clusters where the median radial infall velocity is seen to vary from 200 to 700 km~s$^{-1}$  at 2--4 $R_{200}$ \citep{2021A&A...646A.105P}. Finally, the fact that the diffuse gas in the clouds, arms and bridge in Table~\ref{params}, and in the tail \citep{2008ApJ...682L..85K}, all has velocities which are lower than NGC 4532 by 30--200 km~s$^{-1}$ is consistent with the expectations of ram-pressure stripping. The position-velocity map in Figure~\ref{fig:PV} shows that all the extended gas detected by WALLABY away from NGC 4532 and DDO 137 lies at velocities which are lower by up to 280 km~s$^{-1}$.

\begin{figure}
\begin{center}
\includegraphics[width=1.0\columnwidth,angle=-0]{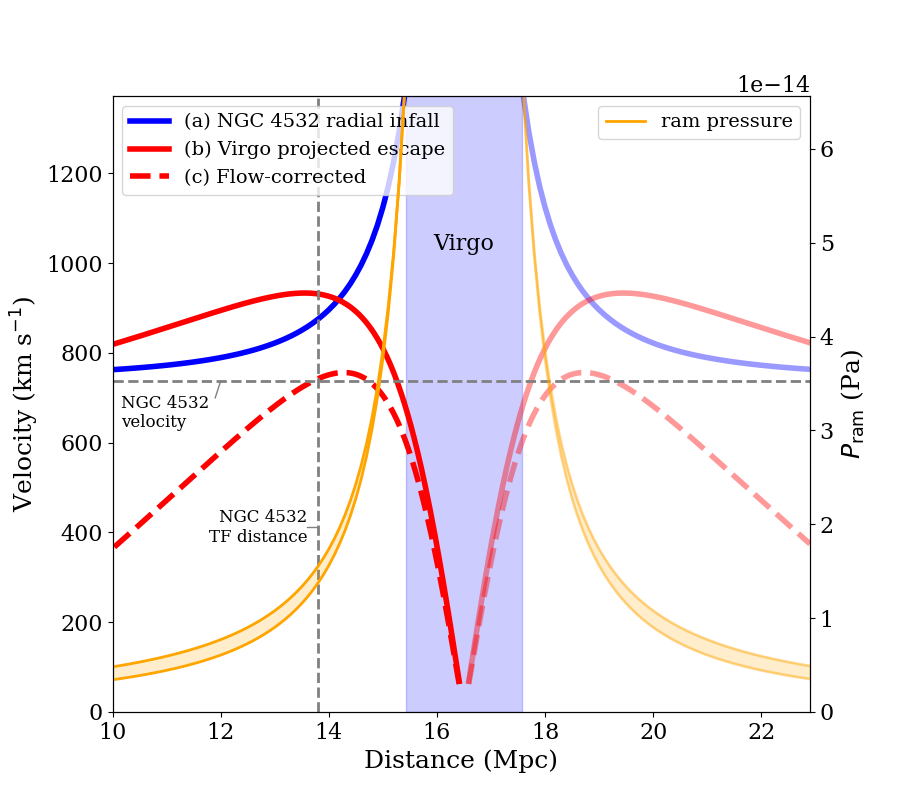}
\caption{A dynamical model of radial infall into the Virgo cluster, and associated ram pressure. The solid blue line (a) represents the deprojected infall velocity of NGC 4532/DDO 137 (WALLABY J123424+062511) into the Virgo cluster as a function of its assumed distance from us (outflow if its distance is greater than Virgo). The blue shaded region represents the distance (and twice the virial radius) of Virgo (this is just for reference since the sightline does not actually go through the virial region). The vertical and horizontal grey dashed lines represent the measured Tully-Fisher distance of NGC 4532, and its radial velocity offset from Virgo, respectively. The solid red line (b) represents the Virgo escape velocity along the sightline towards NGC 4532, corrected for projection. The dashed red line (c) represents a correction for Hubble flow assuming H$_0=70$ km~s$^{-1}$~Mpc$^{-1}$. The orange shaded regions represent the ram pressure in Pa (right-hand vertical axis), calculated from the infall velocity given by the blue line, and the $n_{\rm e}$ density profiles of \citet{2001ApJ...561..708V} and \citet{2011MNRAS.414.2101U} along the NGC 4532 sightline.
}
\label{fig:dynamic}
\end{center}
\end{figure}

A simple dynamical model of radial infall into the Virgo cluster, which takes into account projection, and which assumes that the total mass profile is approximated by the ICM and total mass parameterisations of \citet{2011MNRAS.414.2101U} and \citet{2001ApJ...561..708V} is presented in Figure~\ref{fig:dynamic}. This mass profile is consistent with a mass within the virial radius of 1.08 Mpc \citep{2011MNRAS.414.2101U} of $M_{\rm vir} \approx 1.5\times 10^{14}$ M$_{\odot}$, and a mass within the turnaround radius of 7.3 Mpc \citep{2018ApJ...858...62K} of $M_{\rm ta} \approx 4.4\times 10^{14}$ M$_{\odot}$. 

Three velocity curves are presented: (a) the true de-projected infall velocity of NGC 4532 into Virgo as a function of the assumed distance of NGC 4532, under the assumption of radial motion (the projected velocity difference of 737 km s$^{-1}$ and approximate Tully-Fisher distance of NGC 4532 are also indicated); (b) the radial escape velocity from Virgo based on the assumed mass profile and Virgocentric radius, corrected for projection; and (c) the projected escape velocity along the line of sight, corrected for Hubble flow. The latter correction is a crude approximation given that motions are non-linear inside the turnaround radius. For all distances less than 15.2 Mpc (or greater than 17.8 Mpc, but this would not be consistent with the direction of the tail), the projected NGC 4532 infall velocity (737 km s$^{-1}$) is less than the projected Virgo escape velocity. Even after correction for Hubble flow, the expected escape velocity is consistent with the infall velocity at the NGC 4532 distance. For example, at a Tully-Fisher distance of 13.8 Mpc, the Virgocentric radius of NGC 4532 is 3.2 Mpc, its radial infall velocity is 880 km~s$^{-1}$, the local electron density is $n_{\rm e}\approx 1.2\times10^{-5}$ cm$^{-3}$, and the ram pressure is $1.5\times10^{-14}$ Pa. {The 0.5~Mpc HI tail, if purely radial in extent, would de-project to 0.9 Mpc long, and have a dynamical timescale of $\sim$ 1~Gyr. That is, if the HI in the tail were brought immediately to rest with respect to a static cluster medium, a timescale of $\sim$ 1~Gyr would be required to traverse the deprojected distance of 0.9~Mpc at the deprojected velocity of 880~km~s$^{-1}$.}

\begin{figure}
\begin{center}
\includegraphics[width=1.0\columnwidth,angle=-0]{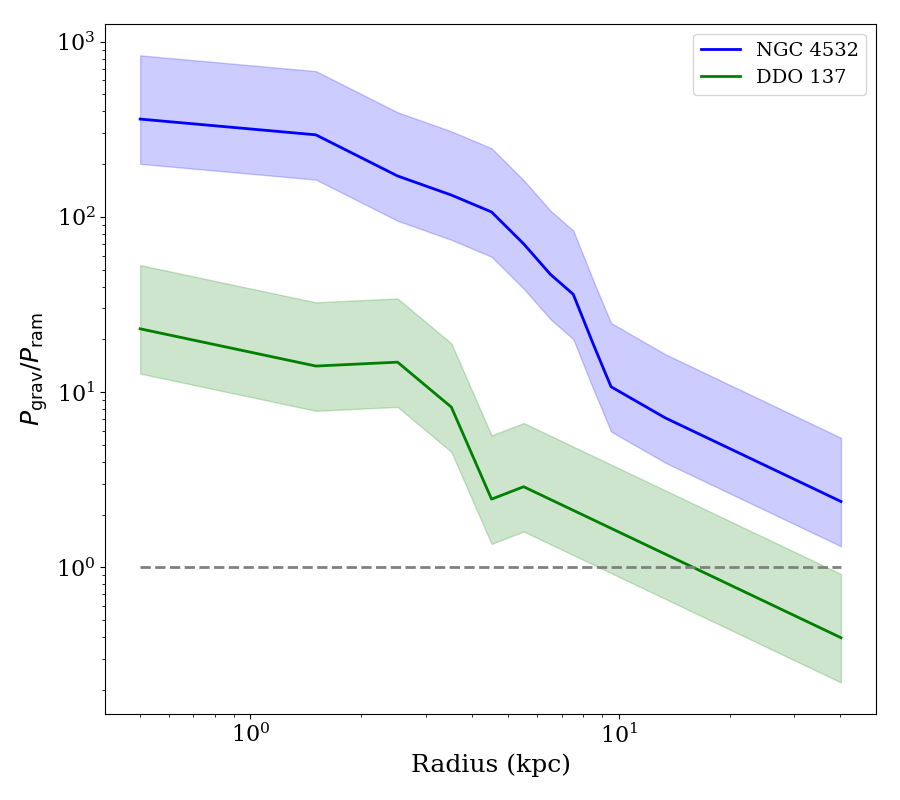}
\caption{The ratio of gravitational restoring pressure to ram pressure for NGC 4532 (upper curve) and DDO 137 (lower curve) based on the rotation curves and surface density distributions presented in Figure~\ref{fig:rotcur}, an ICM density $n_{\rm e} = 1.2\times10^{-5}$ cm$^{-3}$, and a velocity of $880\pm300$ km~s$^{-1}$ (see text for details). Beyond the galaxy disks, the mean HI column density for the bridge of $1.0\times 10^{20}$ cm$^{-2}$ is used, and flat rotation curves are assumed. A distance of 13.8 Mpc is also assumed.
}
\label{fig:ram}
\end{center}
\end{figure}

The consequence for the gas disks of NGC 4532 and DDO 137 of this ram pressure scenario is illustrated in Figure~\ref{fig:ram} where the ratio of gravitational restoring pressure to ram pressure, $(P_{\rm grav}/P_{\rm ram})=(\pi \Sigma_{\rm g} v_{\rm rot}^2) / (n_e v_{\rm ICM}^2 R)$ is plotted against radius $R$ from each galaxy, where $\Sigma_{\rm g}$ is the gas surface density, $v_{\rm rot}$ is the rotation velocity, and $n_{\rm e}$ and $v_{\rm ICM}$ are the ICM density and velocity, respectively \citep{2021A&A...645A.121V}. For the main galaxy disks, the ratio of gravity to ram pressure varies from 10--300, so the disks themselves will be stable until higher values of $n_{\rm e}$ are encountered. However, as shown in Section~\ref{sec:tidal}, the tidal encounter between the two galaxies, in a similar manner to the LMC/SMC system, has likely thrown a sizeable mass of gas from the outer parts of their respective disks (as well as a few stars) to higher points in the potential well of the system, making this gas more susceptible to being stripped away by ram pressure \citep{2015ApJ...813..110H}. Indeed, Figure~\ref{fig:ram} shows that, for positions along the bridge, the gravity/ram pressure ratio is close to unity, even in the unlikely event of the rotation curves of the galaxies remain flat out to 40 kpc. \citet{2001ApJ...563L..23G} have likewise proposed that mutual interaction between CGCG 97-073 and 97-079 in the A1367 cluster has loosened the potential well and contributed to efficacy of ram pressure stripping. Similarly, \citet{2024A&A...690A...4S} propose that NGC 1427A in the Fornax cluster is the product of a high-speed tidal interaction and ram pressure stripping.

Although we have modelled Virgo as having isotropic mass and electron density profiles centred on M87, this is probably not the case much beyond $r_{200}$ where X-ray clumping factor is already seen to rise \citep{2024A&A...689A.113M}. Virgo is a dynamically young system and has a complex morphology as is evident in the X-ray image shown in Figure~\ref{fig:rosat} and maps of galaxy distribution \citep{2018A&A...614A..56B}. For example, the Virgo `cluster B' centred on M49 (see Figure~\ref{fig:rosat}) is estimated to have a mass of around $10^{14}$ M$_{\odot}$ \citep[e.g.][]{2018A&A...614A..56B}. \citet{2024A&A...689A.113M} point to the existence of a 320-kpc long filamentary overdensity southwest of M49 (partially visible in Figure~\ref{fig:rosat}), possible connected with galaxy infall on the far side of Virgo.
Indeed, in projection, the NGC 4532/DDO 137 system is closer to cluster B than the more massive `cluster A' centred on M87. However, according to NED, the mean redshift-independent distance measurement for M49 is 16.1 Mpc, which suggests that the NGC 4532-M49 distance is 2.4 Mpc, which is not much less than the estimated NGC 4532-M87 separation of 3.2 Mpc. It is therefore unlikely that the M49 sub-cluster would itself play much of a role in generating the high infall velocity observed for NGC 4532. However, the role of cluster B and other Virgo sub-clusters, in generating IGM densities through dynamical interactions and stripping of newly infalling galaxies is undoubtedly important in the ensuring efficacy of ram pressure stripping in the region.

\subsection{Combining tidal and ram pressure models}

{As demonstrated in Section~\ref{sec:tidal}, tides cannot be solely responsible for the nature of the NGC 4532/DDO 137 system. Likewise, pre-infall ram pressure cannot be solely responsible for the nature of the system as the available ram pressure is too weak to have displaced enough gas from the disks of the galaxies. }

{However, as seems to be the case for the LMC/SMC system, a combination of the two models fits the data well. The tidal model predicts that the simple gravitational interaction between the binary pair is able (within a Gyr) to displace gas into a bridge, arms and other extended features. Also that rapid merging of the binary pair is suppressed, so that gas remains loosely bound for longer. The loosely-bound gas at these positions ($>10$ kpc from either galaxy) is then less able to resist ram pressure (see Figures~\ref{fig:sim} and \ref{fig:ram}), which is able to strip this gas. The key ingredients for a successful combined model are therefore: (a) the system is within a few Mpc of M87; (b) it is infalling for the first time with a large velocity, and (c) that there is a significant ambient IGM density. These ingredients appear remarkably consistent with the observations.}

Compared with the Magellanic System, which is moving at a velocity of 300 km s$^{-1}$ through a medium of density $\sim 10^{-4}$ cm$^{-3}$ and temperature $10^5$ K \citep{2016ARA&A..54..363D}, the NGC 4532 system appears to be moving at a velocity 2--3 times larger through a medium of density a factor of 10 lower, with a temperature expected to be a factor of 10 higher. The metallicities of the HI tails are however expected to be similar \citep{2003AJ....125.2975L, 2021MNRAS.505.4289P, 2022MNRAS.517.4497D}. 

The lack of UV absorption along the sightline to QSO SDSS J123235.82+060310.0 at the velocity of the HI tail \citep{2016MNRAS.459.1827P} remains a puzzle given its position within the ALFALFA contours. This could be due to structure in the hot gaseous medium \citep{2021MNRAS.506..139M}, the nature of the hot-cold interface \citep{2022MNRAS.511..859G}, or the metallicity, density and ionisation state of the shocked gas \citep{2010ApJ...709.1203T}. We have re-examined the HST archival data (proposal ID: 13383) and interestingly find that the UV spectrum also shows no Ly-$\alpha$ absorption at the tail velocity -- the nearest detected absorption lines are at $-740$ and 570 km~s$^{-1}$ with respect to the tail. Therefore, there is neither cool  nor warm/hot gas present at the position of the QSO. This must be a result of small scale structure in at least one spatial dimension -- possibly filamentary or cloudlet structures akin to that seen in the Magellanic Stream \citep{2002ApJ...576..773S, 2003ApJ...586..170P, 2005A&A...432...45B}. Any `onion skin' of shocked gas must be more moderate in spatial extent than for the Magellanic Stream, implying that radiative cooling and self-shielding mechanisms have been sufficient to have kept the tail mainly neutral at this stage of its evolution, despite being ram pressure stripped. 

In general, there is a large variety in the mass ratio of the cold neutral (HI and H$_2$) and ionised gas phases in the tails of galaxies infalling into clusters. {Whether tails are dominated by warm or cold gas is still unclear, but simulations suggest that this is likely determined by a combination of orbital speed, ISM and ICM temperature and pressure \citep[e.g.][]{2011ApJ...731...98T}.}
The presence of the  dwarf galaxy Tol 1232+052 at the end of the HI tail of WALLABY J123424+062511 supports the presence of gas which is mostly cool and clumpy at this stage. In high-density cluster environments, there are many examples of star-forming dwarf galaxies and 'fireballs' which have likewise formed from a combination of tides and ram pressure \citep{2007MNRAS.376..157C, 2022ApJ...935...51J, 2023ApJ...958...73G}.

We have demonstrated that short range tidal forces could be responsible for the extended features in the vicinity of the system, and that ram pressure could be responsible for the HI tail. Our simulations have also suggested that the long range tidal field from Virgo has acted to suppress binary merging and has therefore enhanced the efficacy of ram pressure stripping and gas removal. It will be interesting to explore the uniqueness and spatial distribution of similar systems in future WALLABY observations. The frequency of such systems  will give insights into the relative importance of ram pressure and tidal forces in different environments. An important tool in this regard will be the deep all-sky observations from eROSITA \citep{2021A&A...647A...1P}, which will be useful in tracing hot gas further in the outskirts of rich groups and clusters of galaxies.

\section{Summary and future work}

WALLABY observations have exposed the full extent of the bridge of neutral gas joining NGC 4532 and DDO 137 (together also known as WALLABY J123424+062511), and highlighted the joint role of ram pressure and the mutual tidal interaction between the two galaxies in explaining the origin of the diffuse gas envelope around the system, and the huge tail previously found by Arecibo. As opposed to the spectacular, but short-lived, ram pressure tails traced by HI, CO, H$\alpha$ and X-ray observations in rich galaxy clusters, the NGC 4532/DDO 137 HI features appear to have remained visible (i.e. neutral) for a longer time owing to the gas either being bound to the galaxy pair or existing within a low density IGM environment. Our models suggest that WALLABY J123424+062511 lies on the near-side of the Virgo cluster, has a radial infall velocity of 880 km s$^{-1}$ and is moving through a region of density $1.2\times 10^{-5}$ cm$^{-3}$. 

The effect of the large-scale tidal influence of the Virgo cluster has been explored, but this cannot explain the length and orientation of the long HI tail for the most plausible geometry. 
Nevertheless, the effects of the Virgo tidal field will assist in the formation of a tail even during the infall period and will delay the merger of the binary pair. Overall the system has many similarities with the Magellanic System, the bridge detected in the WALLABY observations and the huge tail, so further study may assist in our understanding of both.

Future studies could include deeper HI observations of WALLABY J123424+062511 to better map the extent of its diffuse halo, clouds and arms, and their connection to the 0.5 Mpc tail. CO observations would usefully serve to explore in-situ star formation in the bridge, and the possible formation of new dwarf galaxies. An accurate TRGB distance would serve to confirm the validity or otherwise of the infall model.

\section*{Acknowledgements}

{We thank the anonymous referee for suggesting useful improvements}.
We thank Saksham Chopra for his assistance with catalogue cross-matching. 
This scientific work uses data obtained from Inyarrimanha Ilgari Bundara, the CSIRO Murchison Radio-astronomy Observatory. We acknowledge the Wajarri Yamaji People as the Traditional Owners and native title holders of the Observatory site. CSIRO’s ASKAP radio telescope is part of the Australia Telescope National Facility.\footnote{\url{https://ror.org/05qajvd42}} Operation of ASKAP is funded by the Australian Government with support from the National Collaborative Research Infrastructure Strategy. ASKAP uses the resources of the Pawsey Supercomputing Centre. Establishment of ASKAP, the Inyarrimanha Ilgari Bundara, the CSIRO Murchison Radio-astronomy Observatory and the Pawsey Supercomputing Centre are initiatives of the Australian Government, with support from the Government of Western Australia and the Science and Industry Endowment Fund.
{This work was supported by the Australian SKA Regional Centre (AusSRC), Australia’s portion of the international SKA Regional Centre Network (SRCNet), funded by the Australian Government through the Department of Industry, Science, and Resources (DISR; grant SKARC000001). AusSRC is an equal collaboration between CSIRO – Australia’s national science agency, Curtin University, the Pawsey Supercomputing Research Centre, and the University of Western Australia}. The Legacy Surveys imaging of the DESI footprint is supported by the Director, Office of Science, Office of High Energy Physics of the U.S. Department of Energy under Contract No. DE-AC02-05CH1123, by the National Energy Research Scientific Computing Center, a DOE Office of Science User Facility under the same contract; and by the U.S. National Science Foundation, Division of Astronomical Sciences under Contract No. AST-0950945 to NOAO. This research has made use of the NASA/IPAC Extragalactic Database (NED), which is operated by the Jet Propulsion Laboratory, California Institute of Technology, under contract with the National Aeronautics and Space Administration. Parts of this research were supported by the Australian Research Council Centre of Excellence for All Sky Astrophysics in 3 Dimensions (ASTRO 3D), through project CE170100013 and Discovery Project DP210100337 (LC). KS acknowledges funding from the Natural Sciences and Engineering Research Council of Canada (NSERC).

\section*{Data Availability}

The WALLABY source catalogue and associated data products (e.g. cubelets, moment maps, integrated spectra, radial surface density profiles) are available online through the CSIRO ASKAP Science Data Archive (CASDA)\footnote{\url{https://research.csiro.au/casda}} and the Canadian Astronomy Data Centre (CADC).\footnote{\url{https://www.cadc-ccda.hia-iha.nrc-cnrc.gc.ca}} All source and kinematic model data products are mirrored at both locations. Links to the data access services and the software tools used to produce the data products as well as documented instructions and example scripts for accessing the data are available from the WALLABY Data Portal.\footnote{\url{https://wallaby-survey.org/data}}



\bibliographystyle{mnras}
\bibliography{references} 






\bsp	
\label{lastpage}
\end{document}